\documentclass[10pt,journal,compsoc]{IEEEtran}

\usepackage{times}
\usepackage{graphicx}
\usepackage{epstopdf}
\usepackage{epsfig}
\usepackage{amsmath}
\usepackage{subfig}
\usepackage[normalem]{ulem}
\usepackage{setspace}
\usepackage{adjustbox}
\usepackage[noend]{algpseudocode}
\usepackage{algorithmicx,algorithm}
\usepackage{color}
\usepackage{ulem}
\usepackage{url}

\ifCLASSOPTIONcompsoc
  \usepackage[nocompress]{cite}
\else
  \usepackage{cite}
\fi

\ifCLASSINFOpdf
\else
\fi

\begin{document}

\title{IAD: Interaction-Aware Diffusion Framework in Social Networks}
%
%

\author{Xi~Zhang,~\IEEEmembership{Member,~IEEE,}
        Yuan~Su,
        Siyu~Qu,
        Sihong~Xie,~\IEEEmembership{Member,~IEEE,}
        Binxing~Fang,
        and~Philip~S.~Yu,~\IEEEmembership{Fellow,~IEEE}
\IEEEcompsocitemizethanks{\IEEEcompsocthanksitem X. Zhang, Y. Su, S. Qu and B. Fang are with the Key Laboratory of Trustworthy Distributed Computing and Service (Beijing University of Posts and Telecommunications), Ministry of Education, Beijing 100876, China.
\protect\\
E-mail: \{zhangx, timsu, qusiyu, fangbx\}@bupt.edu.cn.
\IEEEcompsocthanksitem S. Xie is with Computer Science and Engineering Department, Lehigh University,
Bethlehem, PA 18015, USA.
\protect\\
E-mail: sxie@cse.lehigh.edu.
\IEEEcompsocthanksitem P.S. Yu is with the Department of Computer Science, University of Illinois
at Chicago, Chicago, IL 60607, USA, and the Institute for Data Science,
Tsinghua University, Beijing 100084, China. 
\protect\\
E-mail: psyu@uic.edu.}
\thanks{Manuscript received April 19, 2005; revised August 26, 2015.}}

%
%

\markboth{Journal of \LaTeX\ Class Files,~Vol.~14, No.~8, August~2015}%
{Shell \MakeLowercase{\textit{et al.}}: Bare Demo of IEEEtran.cls for Computer Society Journals}
%



\IEEEtitleabstractindextext{%
\begin{abstract}
In networks, multiple contagions, such as information and purchasing behaviors, may interact with each other as they spread simultaneously. However, most of the existing information diffusion models are built on the assumption that each individual contagion spreads independently, regardless of their interactions. Gaining insights into such interaction is crucial to understand the contagion adoption behaviors, and thus can make better predictions. In this paper, we study the contagion adoption behavior under a set of interactions, specifically, the interactions among users, contagions' contents and sentiments, which are learned from social network structures and texts. We then develop an effective and efficient interaction-aware diffusion (IAD) framework, incorporating these interactions into a unified model. We also present a generative process to distinguish user roles, a co-training method to determine contagions' categories and a new topic model to obtain topic-specific sentiments. Evaluation on the large-scale Weibo dataset demonstrates that our proposal can learn how different users, contagion categories and sentiments interact with each other efficiently. With these interactions, we can make a more accurate prediction than the state-of-art baselines. Moreover, we can better understand how the interactions influence the propagation process and thus can suggest useful directions for information promotion or suppression in viral marketing. 
\end{abstract}


}

\maketitle

\IEEEdisplaynontitleabstractindextext

%
\IEEEpeerreviewmaketitle

\IEEEraisesectionheading{\section{Introduction}\label{sec:introduction}}

\IEEEPARstart{S}ocial networks are the fundamental medium for the diffusion of information. This diffusion process imitates the spread of infectious disease. Specifically, when a user forwards a contagion (such as a political opinion or product), infection occurs and the contagion gets spread along the edges of the underlying network. The process of a user examining the contagions shared by their neighbours and forwarding some of them results in the information cascades. For the cascade of one contagion, both the users on the cascade and their neighbours are exposed to the contagion, and their behaviors and decisions will be influenced not only by the contents of the contagion but also by their social contacts. In order to effectively employ information diffusion for viral marketing, it is essential to understand the users' adoption decisions under such influences.

%
%
%

Much work has been done to understand the information dynamics in social networks, including theoretical models and empirical studies. However, most of the existing studies assume that each piece of information spreads independently~\cite{Maximizing,Talk,mathematics,structure,Scalable,Sketch,martingale}, regardless of the interactions between contagions. In the real world, multiple contagions may compete or cooperate with each other when they spread at the same time. For example, the news about the banning of Samsung Galaxy S7 in airports may promote the spreading of the battery explosion events of S7, but suppress the news that Samsung is releasing other exciting products. Thus, in this example, the contagion-contagion interaction can be seen as a ``competition" between the popularity of two pieces of information. Taking the interactions into account is crucial to address the question of how much a user would like to adopt a contagion. Recent diffusion models have started to consider interactions between contagions~\cite{Competition,Clash,InnoNetwork,CSI,Quickmeme,Products,Generalized,Winner,Competing}, however, in most cases, the interactions they learned are latent factors and thus are difficult to understand. For example, in~\cite{Clash}, the interactions it considered are between latent topics, making the promotion or suppression effects difficult to interpret. Specifically, given two contagions that are of unrelated content or subject matter, it is difficult to infer whether they will interact with each other when they spread simultaneously.  Actually, what interests us is the interactions among explicit categories, namely whether contagions belonging to one category (say food) would have some positive/negative effects on the spreading of contagions belonging to another category (say health). These interactions can be used to design viral marketing strategies to promote or suppress some products or news. For example, if it can be inferred that contagions belonging to sports usually have positive effects on the adoption of energy drinks, advertisements on energy drinks can be exhibited alongside with sports news in a user's input stream of posts to promote the sales. However, it is challenging to assign each contagion to its category by human due to the large volume in social networks. How to find an efficient way to classify contagions with only minimum supervision is one of the key challenges in this work. Even if a methodology is proposed to obtain explicit categories, the starting point is always a set of latent topics. 

In addition to the categories that each contagion belongs to, recent studies show that the sentiments exhibited in the contagions may also affect the dynamics of information diffusion~\cite{fan2014anger,ferrara2015quantifying}. Suppose that some user has just been infected by negative news (e.g, disasters or losing games) and thus in bad mood, and when she is exposed to a funny story, it is less likely for her to forward it. On the contrary, in happy days (e.g., holidays), users are more likely to forward positive news than negative news. Thus, sentiment analysis is crucial for information diffusion as well and can provide a new dimension to understand the users' adopting behaviours. However, how to effectively uncover the sentiments from contagions remains an open problem. Furthermore, as the sentiments and topics are both extracted from the contagions, they are hence not independent of each other and their coupling relations should also be taken into account in the information diffusion process, which has not been studied in previous works.

Apart from the categories and sentiments extracted from the contagions, social roles have also been proved to play important roles in the information diffusion process~\cite{Role}. Besides, a user may play multiple roles with respect to different communities, and each social role may present different influences on their neighbours. To study how social roles affect the diffusion process, it is necessary to distinguish the users' social roles and then capture their interactions when modeling the user's adoption decisions.

Since there are interactions among contagions and the interactions among users in the information diffusion process, it is natural to ask whether there are interactions between users and contagions. The answer is obvious since most of the users have their own preferences on some messages, and they would like to forward those to their tastes and skip the others. Although a few methods have been proposed to estimate a user's taste on a specific topic, it remains an open question from the perspective of information diffusion, especially when complex interactions are involved.

%

\begin{figure}
\centering
\includegraphics[width=\linewidth]{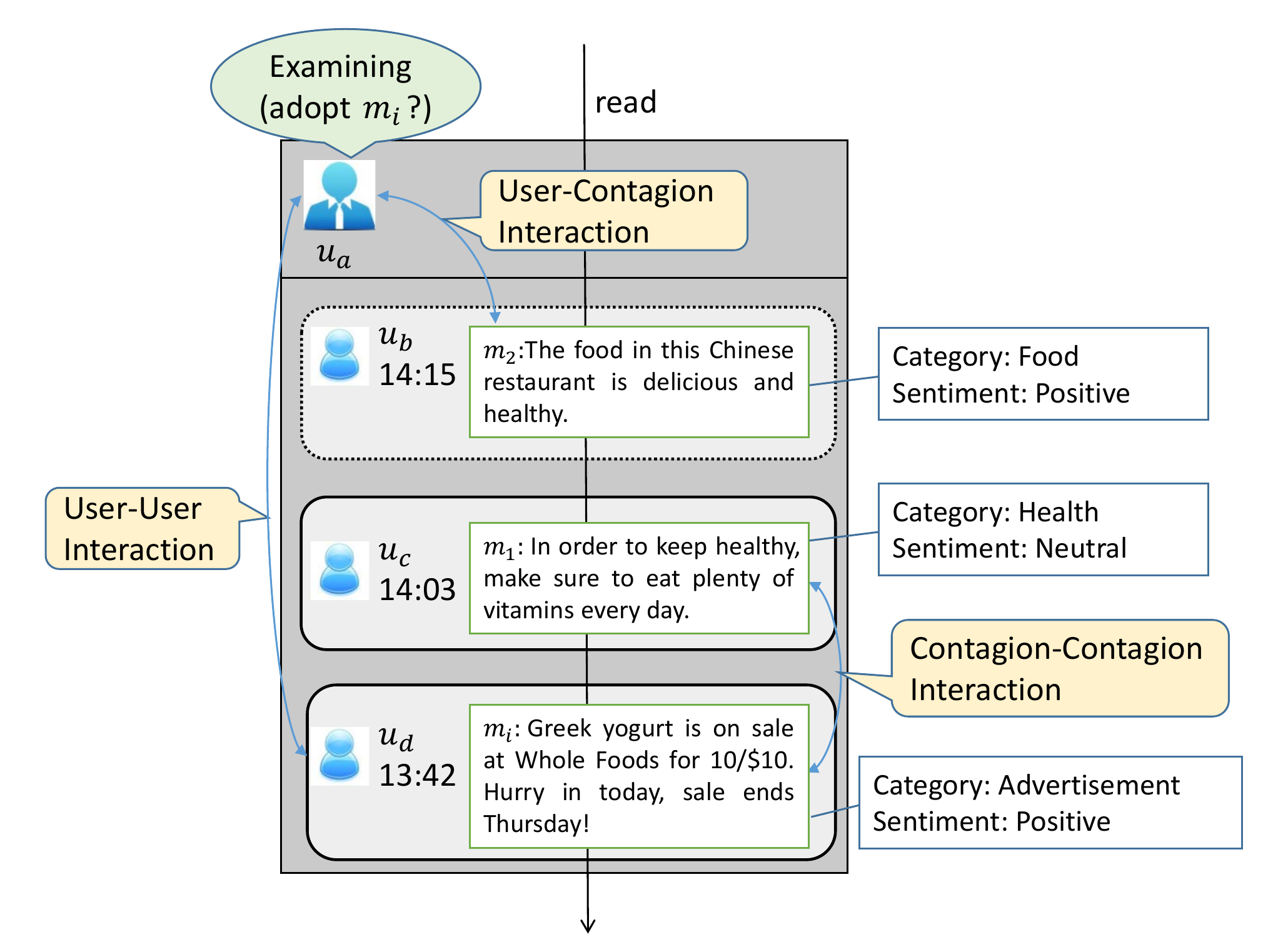}
\caption{An example of the interacting scenario. User $u_a$ is exposed to contagions \{$m_1$, ..., $m_K$\} (here $K=2$) and $m_i$ (forwarded by $u_a$'s neighbour $u_b$), and is examining whether to adopt $m_i$.  $u_a$ 's decision is influenced by the interaction between $u_a$ and $m_i$, the interaction between $u_a$ and $u_d$, and interactions among $m_i$ and other exposing contagions ($m_1$ and $m_2$).}
\label{problem}
\end{figure}

Motivated by the above examples and prospects, there is a clear opportunity to extend the present understanding of information diffusion by jointly studying the interactions existed among the contagions and users. To this end, this paper focuses on the discovery of the interactions of users and contagions in the information diffusion process, as well as the impact of these interactions on the user's adoption behavior. The scenario we study is that when a user is exposed to a set of contagions posted or forwarded by her neighbours, whether she would like to forward a specific one of them. The task we conduct is hence to predict the infection probability by exploiting the inherent popularity of the contagion as well as the interactions among contagions and users. We approach this problem through an interaction-aware diffusion model (IAD), which considers three kinds of interactions: (1) \emph{User-Contagion Interaction}, (2) \emph{User-User Interaction}, (3) \emph{Contagion-Contagion Interaction}. Fig.~\ref{problem} describes an example of the three kinds of interactions when user $u_a$ is exposed to a set of contagions posted by her neighbours. When $u_a$ is examining whether to adopt the contagion $m_i$ posted by $u_d$, there exists \emph{User-User interactions} between $u_a$ and $u_d$, \emph{User-Contagion Interaction} between $u_a$ and $m_i$, as well as \emph{Contagion-Contagion Interaction} between $m_1$ and $m_i$. We then describe each contagion from two views, say category and sentiment, and refine the contagion-related interactions and capture \emph{Category-Category Interaction}, \emph{Sentiment-Sentiment Interaction}, \emph{User-Category Interaction}, and \emph{User-Sentiment Interaction}. As the interactions are rather complex and often exhibit strong coupling relations, rather than exploiting each kind of the interactions separately, we incorporate all the interactions into a unified framework.

Due to the large volume of contagions and users in the online social networks, learning interactions for each pair of contagions and users is impractical. To address this challenge, we use the interactions among the categories of users and contagions instead. Specifically, we first apply a mixture of Gaussians model to explain the generation process of user network features, and use EM algorithm to extract the social role distribution for each user. We then propose a new topic model to capture topics and sentiments simultaneously for each contagion. To get the category of each contagion, a classification approach based on co-training~\cite{co-training} is then developed, with only a small number of labeled data. Overall, our model can statistically learn the interactions effectively, aiming to better understand the information diffusion process and provide a more predictive diffusion model.

Our contributions are summarized as follows:

\begin{itemize}
\item[1)]  We propose an IAD framework to model the user's forwarding behavior by jointly incorporating various kinds of interactions: interactions among contagions, interaction among users as well as interactions among users and contagions. This framework provides new insights into how forwarding decisions are made.
    \\
\item[2)] Due to the large volume of interactions to learn, we devise a co-training based method to classify the contagions to categories, and apply a generative process to obtain the social roles for users, to significantly reduce the complexity of the fitting process.
    \\
\item[3)] To obtain the sentiment-related interactions from contagions, we develop a new topic model called Latent Dirichlet Allocation with Sentiment (LDA-S), a variation of LDA model~\cite{LDA}, to obtain the sentiment and topic simultaneously in short texts.
    \\
\item[4)] Experimental evaluation on a large-scale Weibo dataset~\cite{weibodataset} shows that IAD outperforms state-of-art works in terms of F1-score, accuracy and fitting time. Besides, the estimated interactions reveal some interesting findings that can provide compelling principles to govern content interaction in information diffusion.

\end{itemize}

The remainder of this paper is organized as follows. Section 2 introduces the related works. In Section 3, we describe the proposed IAD framework. Section 4 describes how to classify contagions to explicit categories and Section 5 illustrates the process to infer sentiment-related interactions. In Section 6, extensive experiments have been conducted to show the effectiveness of the proposed approach. Section 7 concludes this paper.

\section{Related Work}

\subsection{Information Diffusion}

In recent years, researchers have extensively studied the information diffusion in social networks~\cite{7050284}. A collection of models are proposed to explain the diffusion process from various perspectives \cite{Inferring,Cascading,Who,Everyone,Minimal,Modeling,zhang2016near,7921565}, while some other models are proposed to predict whether a piece of information will diffuse \cite{PredictMessage,Clash}.

However, most of the prior models assume the spreading of each piece of information is independent of others, e.g., the Linear Threshold Model \cite{Threshold,Micromotives}, the Independent Cascade Model \cite{Maximizing,Talk}, SIR and SIS Model \cite{mathematics,structure}. The diffusion of multiple contagions has been covered in several recent works \cite{Quickmeme,Products,Generalized,Winner,Competing}. The scenario discussed by these works is that one contagion is mutually exclusive to others, i.e., only involving the competition of contagions. In \cite{Competition}, an agent-based model is employed to study whether the competition of information for user's finite attention may affect the popularity of different contagions, but this model does not quantify the interactions between them. Our research is not limited to the mutual exclusivity condition, but instead a user can adopt multiple contagions. We offer a comprehensive consideration for the inter-relations of the contagions in online social networks.

The most related work to ours is the IMM model proposed in \cite{Clash}, which statistically learns how different contagions interact with each other through the Twitter dataset. It models the probability of a user's adoption of information as a function of the exposure sequence, together with the membership of each contagion to a cluster. However, this model doesn't consider the user roles, which has been proved to play an important role in information diffusion in our work. In addition, the clusters in this model are latent variables without real-world meanings. In contrast, our proposal can infer interactions among explicit categories, which are easy to interpret. Nonetheless, IMM model is implemented as a baseline to compare with. Other studies \cite{InnoNetwork,CSI,su2016understanding} also neglect the influence of user roles' interactions, and do not discover the interactions of actual categories of contagions, which are different from our work.

\subsection{Sentiment Analysis}

Although recent works suggest the sentiments in the contents can play important roles in various applications such as product and restaurant reviews~\cite{sent4}, stock market prediction~\cite{sent5}, few existing studies quantify the effects of contents sentiment on the dynamics of information diffusion. Empirical analysis on German political blogosphere indicates that people tend to participate more in emotionally-charged (either positive or negative) discussions~\cite{dang2012impact}. A recent study on Twitter exhibits the effect of sentiment on information diffusion~\cite{ferrara2015quantifying}, and reveals different diffusion patterns for positive and negative messages respectively. However, different from our IAD model, these works still treat each contagion in isolation and thus do not take the interactions into account.

Besides, most of the previous studies try to extract only the sentiments. However, sentiments polarities are often dependent on topics or aspects. Therefore, detecting on which topics of the users are expressing their opinions is very important. Several models have been proposed to infer the topic and sentiment simultaneously. Mei et al. \cite{sent7} propose the TSM model which can reveal the latent topical facets in a Weblog collection, the subtopics in the results of an ad hoc query, and their associated sentiments. Lin et al. \cite{sent4,5710933} propose a novel probabilistic modeling framework based on LDA, called joint sentiment/topic model (JST), which detects sentiment and topic simultaneously from a text. This model assumes that each word is generated from a joint topic and sentiment distribution, and hence doesn't distinguish the topic word and opinion word distributions. Liu et al.~\cite{6990628} propose a topic-adaptive sentiment classification model which extracts text and non-text features from twitters as two views for co-training. Tan et al. ~\cite{tan2014interpreting} propose a LDA based model, Foreground and Background LDA (FB-LDA), to distill foreground topics and filter out longstanding background topics, which can give potential interpretations of the sentiment variations. There are some other topic models considering aspect-specific opinion words~\cite{zhao2010jointly,ho2011exploiting,wang2016mining,schouten2016survey}. A recently proposed TSLDA model can estimate different opinion word distributions for individual sentiment categories for each topic~\cite{sent5}, and has been successfully applied to stock prediction. One weakness of TSLDA is that it divides a document into several sentences and sample the topic and sentiment of each sentence. Therefore, its performance is limited when it is applied to Weibo where most of the messages have only one or two sentences. The other weakness is that it lacks prior information, making it difficult to achieve good results for short texts. To address the aforementioned problems, we propose a variation of TSLDA model, namely LDA-S, to make it work for short texts such as Weibo and Twitter.


\section{Interaction-Aware Diffusion Framework}

In this section, we first state and formulate the problem, and then propose our framework and the corresponding learning process. Before going into details of IAD framework, we define some important notations shown in Table~\ref{Notations}.

%

\begin{table}
\caption{Description of Symbols in IAD Framework} \label{Notations}
\small
\centering
\begin{tabular}{l l} \hline
Symbol &Description\\ \hline
$u$ & Users \\
$m$ & Contagions \\
$r$ & User roles\\
$t$ & Contagion latent topics\\
$s$ & Contagion sentiments\\
$c$ & Contagion categories\\
$\Delta \in R ^ {|u| \times |u|}$ & User-user interaction matrix\\
$\Lambda \in R ^ {|m| \times |m|}$ & Contagion-contagion interaction matrix\\
$\Omega \in R ^ {|u| \times |m|}$ & User-contagion interaction matrix\\
$\Delta_{role} \in R ^ {|r| \times |r|}$ & User role-role interaction matrix\\
$\Lambda_{topic} \in R ^ {|t| \times |t|}$ & Contagion topic-topic interaction matrix\\
$\Lambda_{t-s} \in R ^ {|ts| \times |ts|}$  & Contagion (topic-sentiment) - \\
& (topic-sentiment) interaction matrix\\
$\Lambda_{s} \in R ^ {|s| \times |s|}$  & Contagion sentiment - \\
& sentiment interaction matrix\\
$\Omega_{topic}^{role} \in R ^ {|r| \times |t|}$ & User role - contagion topic\\
&  interaction matrix\\
$\Omega_{t-s}^{role} \in R ^ {|r| \times |ts|}$ & User role - contagion\\
& (topic-sentiment) interaction matrix\\
$\Omega_{s}^{role} \in R ^ {|r| \times |s|}$ & User role - contagion sentiment\\
& interaction matrix\\
$\Lambda_{categ.} \in R ^ {|c| \times |c|}$ & Contagion category-category \\
& interaction matrix\\
$\Omega_{categ.}^{role} \in R ^ {|r| \times |c|}$ & User role - contagion category \\
& interaction matrix\\
\hline
\end{tabular}
\end{table}

\subsection{Problem Statement}\label{sec:model}

In a social network, when some new contagion is originated from a user, we assume that the user's neighbours would see this contagion, or would be exposed to this contagion.
This assumption is consistent with~\cite{Clash}. The exposed contagion is called an exposure. Since users have limited attention~\cite{Competition}, we make the assumption as~\cite{Clash} that at a given time, a user can read through all the contagions her neighbours have forwarded, but only the most recent $K$ exposures that she can keep in mind. In social networks like Weibo and Twitter, tweets in a user's reading screen are arranged in time descending order, i.e., users will first read the most recent contagions and then go backward. \footnote{Please note that the dataset that we analyze was collected in 2012, and at that time Weibo and Twitter still showed tweets in the reversed chronological order. Though they have stopped showing contagions in this simple order at present, our model can still work if we could identify which set of contagions are read simultaneously by a user.} Therefore, there is a sliding attention window going back $K$ contagions, and the $K$ contagions in the window may affect a user's adoption behavior. Then the problem we focus is that when a user reads a contagion that has been forwarded by one of her neighbours, given the sequence of contagions the user has previously read, what's the probability of the user adopting this contagion.

Figure~\ref{problem} describes the interaction scenario studied in this paper, where the set \{$m_1$,$m_2$,...$m_K$\} is a sequence of $K$  contagions user $u_a$ has read and kept in mind, and $m_i$ ($i\neq 1,2,...,K$) is the contagion which is previously forwarded by a user $u_d$ and now examined by $u_a$. $u_a$ will determine whether to adopt (i.e., forward) $m_i$. In this scenario, the forwarding decision made by $u_a$ is not only decided by the inherent characteristics of $m_i$, but also by external interactions described as follows:

\begin{itemize}
\item \emph{{User-Contagion Interaction}}: The interaction between the examining user and the examined contagion. As shown in Fig.~\ref{problem},  it is $u_a$'s preference over $m_i$.
    \\
\item \emph{{User-User Interaction}}: The interaction between the examining user and the neighbour who has forwarded the examined contagion previously. In Fig.~\ref{problem}, it is the effect $u_b$ has on $u_a$.
    \\
\item \emph{{Contagion-Contagion Interaction}}: The interaction among the examined contagion and other contagions the user has read recently. In Fig.~\ref{problem}, it is the effect contagions $m_1$ and $m_2$ ($K=2$) has on $m_i$.
\end{itemize}

Given a collection of the interacting scenarios, our task is to model the users' adoption behaviour by incorporating the aforementioned interactions, and fitting the model to infer the interactions. Meanwhile, we can make more accurate predictions on users' adoption behaviors. The problem will be formulated in the next subsection.

\subsection{Formulation}
According to the interacting scenario, given \{$m_1$,$m_2$,...$m_K$\}  and $u_b$, the probability of infection by $m_i$ to $u_a$ is

\begin{equation}
\small
P(I_{m_i(u_a)}|E_{m_i(u_b)}, E_{\{m_1,m_2,...,m_K\}}),
\label{eqn:goal}
\end{equation}

where $I_{m_i(u_a)}$ is the infection of $u_a$ by $m_i$, $E_{m_i(u_b)}$ is the exposure of $m_i$ which is forwarded by $u_b$, and $E_{\{m_1,m_2,...,m_K\}}$ is the exposure set $\{m_1, m_2,..., m_K\}$. We make the same assumption as~\cite{Clash} that for any $k$ and $l$, $E_{m_k}$ is independent of $E_{m_l}$. Applying Bayes' rule, we model Eq. (\ref{eqn:goal}) by

\begin{equation}
\small
\begin{split}
&P(I_{m_i(u_a)}|E_{m_i(u_b)}, E_{\{m_1,m_2,...,m_K\}}) \\
=&\frac{P(I_{m_i(u_a)})  P(E_{m_i(u_b)}, E_{\{m_1,m_2,...,m_K\}}|I_{m_i(u_a)})}{P(E_{m_i(u_b)}, E_{\{m_1,m_2,...,m_K\}})} \\
=&\frac{P(I_{m_i(u_a)})  P(E_{m_i(u_b)}|I_{m_i(u_a)})  \prod_{k=1}^{K} P(E_{m_k}|I_{m_i(u_a)})}{P(E_{m_i(u_b)})  \prod_{k=1}^{K} P(E_{m_k})} \\
=&\frac{P(I_{m_i(u_a)})  \frac{P(I_{m_i(u_a)}|E_{m_i(u_b)})  P(E_{m_i(u_b)})}{P(I_{m_i(u_a)})}  }{P(E_{m_i(u_b)})  \prod_{k=1}^{K} P(E_{m_k})} \\
& \times \prod_{k=1}^{K} \frac{P(I_{m_i(u_a)}|E_{m_k})  P(E_{m_k})}{P(I_{m_i(u_a)})} \\
=&\frac{P(I_{m_i(u_a)}|E_{m_i(u_b)})}{P(I_{m_i(u_a)})^K} \prod_{k=1}^{K} P(I_{m_i(u_a)}|E_{m_k}).
\label{eqn:bayes}
\end{split}
\end{equation}

Here we need to model $P(I_{m_i(u_a)})$, $P(I_{m_i(u_a)}|E_{m_i(u_b)})$ and $P(I_{m_i(u_a)}|E_{m_k})$ for each $k \in \{1,...,K\}$, which are enforced between $0$ and $1$. Since each contagion has its inherent infectiousness, $P(I_{m_i})$ is defined as the prior infection probability of $m_i$, which can be obtained through dividing the number of its infections by the number of its exposures.

We define $\Omega(u_a,m_i)$ as the effect user $u_a$ has on contagion $m_i$ (\emph{User-Contagion Interaction}), $\Delta(u_a,u_b)$ as the effect user $u_b$ has on user $u_a$ (\emph{User-User Interaction}), and $\Lambda(m_i, m_k)$ as the effect contagion $m_k$ has on contagion $m_i$ (\emph{Contagion-Contagion Interaction}). Then we model $P(I_{m_i(u_a)})$, $P(I_{m_i(u_a)}|E_{m_i(u_b)}) $ and $P(I_{m_i(u_a)}|E_{m_k})$ as

\begin{equation}
\small
P(I_{m_i(u_a)}) \approx P(I_{m_i}) + \Omega(u_a,m_i)
\label{eqn:u-c-inter}
\end{equation}

\begin{equation}
\small
\begin{split}
P(I_{m_i(u_a)}| & E_{m_i(u_b)})  \approx P(I_{m_i(u_a)}) + \Delta(u_a,u_b) \\
& \approx P(I_{m_i}) + \Omega(u_a,m_i)+ \Delta(u_a,u_b)
\label{eqn:u-inter}
\end{split}
\end{equation}

\begin{equation}
\small
\begin{split}
P(I_{m_i(u_a)}| & E_{m_k}) \approx P(I_{m_i}|E_{m_k}) + \Omega(u_a,m_i) \\
& \approx P(I_{m_i}) + \Lambda(m_i, m_k) + \Omega(u_a,m_i)
\label{eqn:c-inter}
\end{split}
\end{equation}

\emph{Example 1}. In Fig.~\ref{problem}, assuming that  $P(I_{m_i})=0.5$, $\Omega(u_a,m_i)=0.02$, $\Delta(u_a,u_b)=-0.03$, $\Lambda(m_i, m_{k1})=-0.04$ and $\Lambda(m_i, m_{k2})=-0.05$, then we can derive that $P(I_{m_i(u_a)}) \approx 0.52$ according to Eq. (\ref{eqn:u-c-inter}), $ P(I_{m_i(u_a)}|  E_{m_i(u_b)})  \approx 0.49$ according to Eq. (\ref{eqn:u-inter}) and $P(I_{m_i(u_a)}|  E_{m_{k1}}) \approx 0.43$ according to Eq. (\ref{eqn:c-inter}). Integrating these into Eq. (\ref{eqn:bayes}), we can obtain that the probability of $u_i$ adopting $m_i$ is 0.327.

Please note that, as shown in  Eq. (\ref{eqn:u-c-inter}), Eq. (\ref{eqn:u-inter}), and Eq. (\ref{eqn:c-inter}), the proposed model adopts summations to combine the interaction matrices to the prior infection probability $P(I_{m_i})$. In addition, we also develop a model which adopts multiplications to combine them, however, the experimental results show that the model with summations performs better. The possible reason is that, the form of Eq. (\ref{eqn:bayes}) will be like a linear function when adopting multiplications, and its expressive power would be weaker compared to that with the additive schema. Thus we apply the additive model in this paper.

Though we have connected the infection probability with three interaction matrices: (1)  $\Omega \in R^{|u| \times |m|}$, (2) $\Delta \in R^{|u| \times |u|}$, and (3) $\Lambda \in R^{|m| \times |m|}$, where $|u|$ is the number of users and $|m|$ is the number of contagions, these matrices are impractical to learn because $|u|$ and $|m|$ are extremely large in social networks. Thus, to decrease the parameters to fit, we model \emph{User Role - Contagion Topic Interaction}, \emph{User Role-Role Interaction} and \emph{Contagion Topic-Topic Interaction} instead. Moreover, we also involve sentiments into IAD framework and model the \emph{User Role - Contagion Sentiment Interaction}. It will be described in detail in the next subsection.

\subsection{The Proposed Approach}\label{approach}

\noindent\textbf{Overview}

To decrease the fitted parameters in the interaction matrices, user roles and contagion categories are introduced, and then the interactions between user roles and categories can be learned efficiently. To this end, we utilize the network structures to infer users' social roles, and use the contagion texts to extract the contagions' topics.
To obtain the sentiments and topics from contagions, we propose an extension of LDA model called LDA-S to extract sentiments. The whole process of IAD framework is shown in Fig.~\ref{framework}, comprising of the following five components:

\begin{figure*}
\centering
\epsfig{file=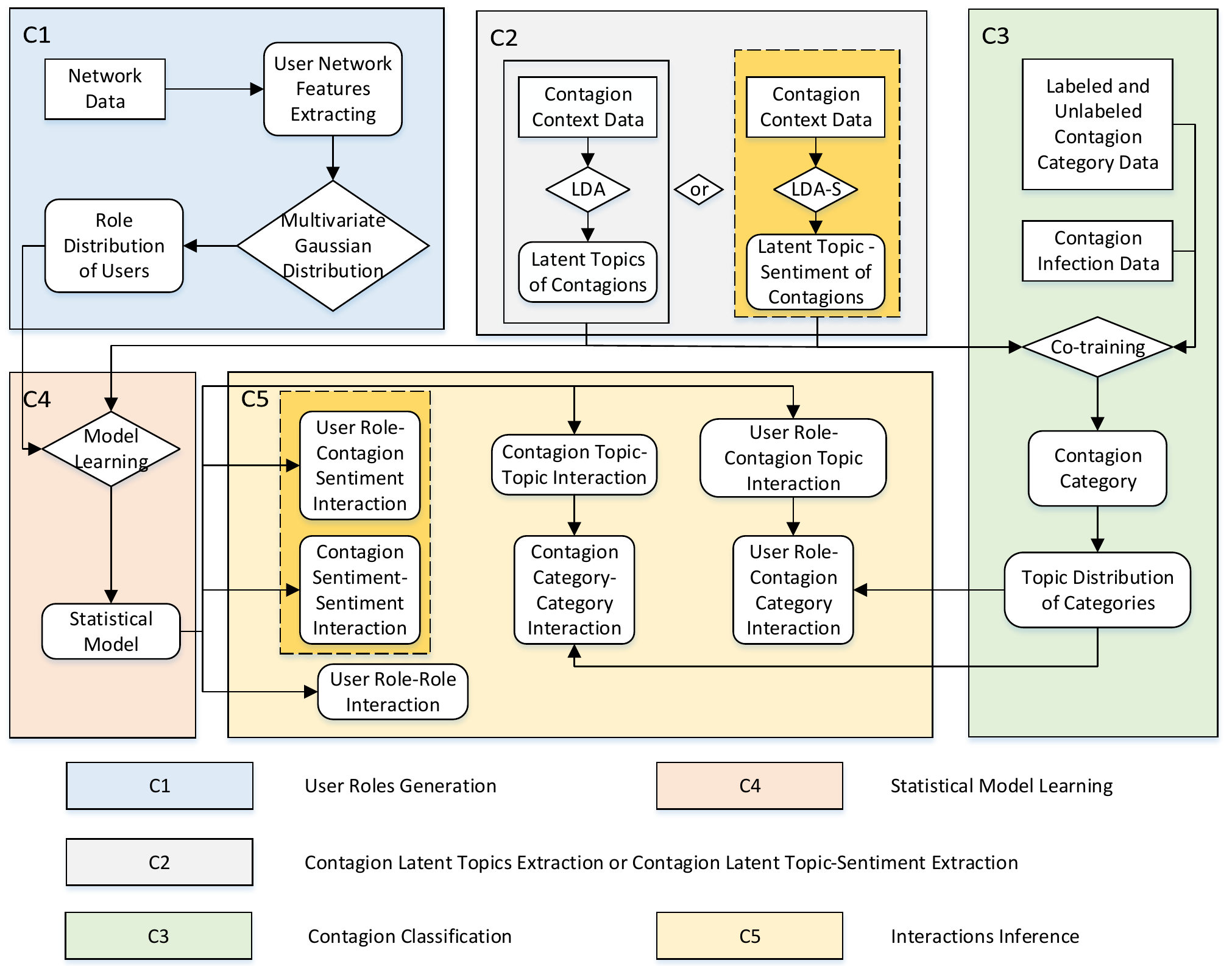, height=4.3in, width=5.6in}
\caption{IAD framework. }
\label{framework}
\end{figure*}

\begin{itemize}
\item \emph{User roles generation }(C1): A generative process for user roles is proposed to distinguish different kinds of users.
    \\
\item \emph{Contagion latent topics extraction or Contagion Latent Topic-Sentiment Extraction} (C2): We have two ways to deal with the contagion texts. One way is to apply LDA model to extract latent topics, and the other is to extract both topics and sentiments with our proposed LDA-S. The output of this component will be used as features for statistical model learning (C4) and contagion classification (C3).
\\

\item \emph{Contagion classification} (C3): Based on the latent topics from C2, a co-training method of contagion classification is proposed to assign the contagions with explicit categories.
    \\
\item \emph{Statistical model learning} (C4): Based on the outputs of C1 and C2, a statistical model is learned.
    \\

\item \emph{Interactions inference} (C5): Given the results of contagion classification (C3) and the statistical model (C4), interactions among contagions and users can be inferred.
\end{itemize}

Next we will introduce the process of user roles generation and contagion latent topic extraction in detail, and then describe statistical model learning. C3 will be illustrated in Section~\ref{sec:classify}. Note that in C2 and C5, the dashed boxes are newly added parts over the earlier version of this work~\cite{suunderstanding}, in order to involve sentiments into the IAD framework, which will be illustrated in Section~\ref{sec:ldas}.

\noindent\textbf{User Role-Role Interaction.}\ \ \
User roles are categorized into \emph{authority users}, \emph{hub users} and \emph{ordinary users} in our work. Intuitively, an \emph{authority user} commonly has a large number of followers, indicating that the node of \emph{authority user} has a large in-degree but its out-degree is small, while a \emph{hub user} has lots of followees, which means the node of \emph{hub user} has a small in-degree but large out-degree. And an \emph{ordinary user} usually does not have a lot of followers or followees, i.e., the in-degree and out-degree of the node are both small. A user may play multiple roles, for instance, an \emph{authority user} may also be a \emph{hub user} to some extent, and therefore we adopt a probability distribution over social roles for each user. Then we infer the interactions among different social roles. The results can demonstrate how a user,  with a specific roles distribution, influence other user's probability of adopting a contagion.

We use PageRank score~\cite{pageRank}, HITS authority and hub values~\cite{HITS}, in-degree and out-degree scores as features of users. A mixture of Gaussians model is proposed to explain the features generation process. Specifically, we assume the features of each user is sampled as a multivariate Gaussian distribution. Intuitively, users with the same roles have similar features and share the same multivariate Gaussian distribution, e.g., two \emph{authority users} are both likely to have a large number of followers. Define \emph{\textbf{r}} = $[r_1,r_2,r_3]^\top$ as a user role vector, representing a probability distribution over social roles for each user, e.g., $[0.5,0.2,0.3]^\top$. Then for each role $r_j$, we generate multivariate Gaussian distribution $u|r_j \sim N(\mu_j, \Sigma_j)$. EM algorithm is used to extract the role distribution for each user. After that, we determine each $r_j$ as the most relevant one of the three roles, according to the fact that authority users commonly have lots of followers and hub users have lots of followees.

Rather than modeling the \emph{User-User Interaction} denoted by $\Delta \in R^{|u| \times |u|}$, we would model \emph{User Role-Role Interaction} instead, which is denoted by $\Delta_{role} \in R^{|r| \times |r|}$. $\Delta_{role}(r_i,r_j)$ is the effect role $r_j$ has on role $r_i$. Define $ \vartheta_{a,i}$ as the probability of user $u_a$ belonging to role $r_i$, and $\sum_{i}\vartheta_{a,i}=1$, and then $ \Delta(u_a, u_b)$  in Eq. (\ref{eqn:u-inter}) can be updated by
\begin{equation}
\small
\Delta(u_a, u_b) = \sum_i\sum_j\vartheta_{a,i}\Delta_{role}(r_i,r_j)\vartheta_{b,j}
\label{eqn:role-inter}
\end{equation}

\noindent\textbf{Contagion Topic-Topic Interaction.}\ \ \
Each contagion is assumed to have a distribution on several topics, and $t$ denotes the set of latent topics. LDA~\cite{LDA} is used to extract the latent topic distribution of each contagion. Then, instead of modeling $\Lambda \in R^{|m| \times |m|}$, we would model a matrix $\Lambda_{topic} \in R^{ |t| \times |t|}$, which denotes the \emph{Contagion Topic-Topic Interaction}. We define $ \theta_{i,a}$ as the probability of contagion $m_i$ belonging to topic $t_a$, and therefore $\sum_{a}\theta_{i,a}=1$.  Let $\Lambda_{topic}(t_a,t_b)$ denote the impact of topic $t_b$ has on topic $t_a$. Now, $ \Lambda(m_i, m_k)$  in Eq. (\ref{eqn:c-inter}) can be updated by

\begin{equation}
\small
\Lambda(m_i, m_k) = \sum_a\sum_b\theta_{i,a}\Lambda_{topic}(t_a,t_b)\theta_{k,b}
\label{eqn:topic-inter}
\end{equation}

Besides the \emph{Contagion Topic-Topic Interaction}, we also propose a topic-sentiment model to get \emph{Contagion (Topic-Sentiment)-(Topic-Sentiment) Interaction} to update Eq. (\ref{eqn:c-inter}).

\noindent\textbf{User Role - Contagion Topic Interaction.}\ \ \
Instead of learning $\Omega$, we build a matrix $\Omega_{topic}^{role} \in R^{ |r| \times |t|}$ to denote the \emph{User Role - Contagion Topic Interaction}. Then $\Omega(u_a,m_i)$ in Eq. (\ref{eqn:u-c-inter}), Eq. (\ref{eqn:u-inter}) and Eq. (\ref{eqn:c-inter}) can be updated by

\begin{equation}
\small
\Omega(u_a,m_i) = \sum_j\sum_b\vartheta_{a,j}\Omega_{topic}^{role}(r_j,t_b)\theta_{i,b}.
\label{eqn:role-topic-inter}
\end{equation}

\subsection{Model Learning}

The input of our model is a set of interacting scenarios. An example of the interacting scenario is shown in Fig.~\ref{problem}, which consists of the examining user $u_a$, the examined contagion $m_i$, user $u_a$'s neighbour $u_b$ who has forwarded the examined contagion, and the exposing contagion set \{$m_1$, $m_2$, ..., $m_K$\} ($i\neq 1,2,...,K$). All the interacting scenarios comprise a set \{$x_1$, $x_2$, ..., $x_n$\}, where $x_i$ is the $i$th interacting scenario and $n$ is the total number of scenarios. For each interacting scenario, we can observe whether the examining user has adopted the examined contagion or not, and denote it as $y_i \in \{0,1\}$ ($1$ for adoption and $0$ for not). Then the training set $\{(x_1,y_1), (x_2,y_2), ..., (x_n,y_n)\}$ will be obtained. Let $\pi(x_i)$ denote Eq. (\ref{eqn:goal}) for simplicity. Now $\pi(x_i)$ can be updated by $\Omega_{topic}^{role}$, $\Delta_{role}$ and $\Lambda_{topic}$, and the log-likelihood function is


\begin{equation}
\small
\begin{split}
&L(\Omega_{topic}^{role}, \Delta_{role}, \Lambda_{topic}) \\
= & \sum_{i=1}^n(y_i log \pi(x_i)+(1-y_i)log(1-\pi(x_i)))
\label{eqn:likelihood}
\end{split}
\end{equation}

Our goal is to estimate the parameters in $\Omega_{topic}^{role}$, $\Delta_{role}$ and $\Lambda_{topic}$ to maximize the log-likelihood function. Stochastic gradient ascent is adopted to fit the model. In each iteration in the parameters updating process, if it makes any variable with probability meaning smaller than 0 or larger than 1, we don't conduct any updating in this iteration, and go directly to the next iteration.

\section{Classification of Contagions}\label{sec:classify}

The interaction matrix $\Lambda_{topic}$ and $\Omega_{topic}^{role}$ learned through our model are comprised of latent topics, which are difficult to interpret. In this section, we illustrate how to obtain interactions among explicit categories. We define $|c|=15$ categories based on the Weibo dataset, involving \emph{advertisement}, \emph{constellation}, \emph{culture}, \emph{economy}, \emph{food}, \emph{health}, \emph{history}, \emph{life},  \emph{movie}, \emph{music}, \emph{news}, \emph{politics}, \emph{sports}, \emph{technology} and \emph{traffic}. One contagion on Weibo is always a short text with only a few words, and is mostly affiliated to only one specific class. Thus, the multi-label classification techniques such as~\cite{Rubin2012} are not applied in this paper.

To discover interactions among categories, contagions should be classified into categories first. However, contagions spreading in Weibo~\cite{weibodataset} are not labeled to intrinsic categories. Labeled contagions are extremely expensive to obtain because large human efforts are required. Thus, only a few labeled contagions are available for learning. A classification approach based on co-training~\cite{co-training} is proposed. Co-training is a semi-supervised learning technique that assumes each example is described using two different feature sets (or views) that provide different, complementary information about the instance. It first learns a separate classifier for each view using an initial small set of labeled examples, and then use these classifiers on the unlabeled examples. The most confident predictions are iteratively added to the labeled training data.

Specifically, for our task, each contagion in the dataset is described in two distinct views. One is the contagion itself, and the other is a set of the other contagions posted by the same user. The intuition here is that contagions created from the same user are prone to have the similar category. Then we build two classifiers for two views, and choose the latent topics as the features for each classifier. As described in Section~\ref{approach}, contagion $m_i$'s latent topic distribution, denoted by $ \theta_{i,a}(a \in 1,..|t|$, where $|t|$ denotes the number of latent topics), can be extracted using LDA. We define a contagion set $M_i = \{m_1, m_2,...,m_k\}$ to contain the other contagions created by the same user. The latent topic distribution $ \Theta_{i,a}(a \in 1,..|t|)$ of $M_i$ is obtained by $\frac{\sum_{j=1}^{k}\theta_{j,a}}{k}$. Now, the two classifiers are listed as follows, and LIBSVM~\cite{LIBSVM} is used for multi-class classification.
\begin{itemize}
\item \emph{\textbf{Classifier 1}}: $ \theta_{i,a}(a \in 1,..|t|)$ as features for each contagion $m_i$.
\\
\item \emph{\textbf{Classifier 2}}: $ \Theta_{i,a}(a \in 1,..|t|)$ as features for each contagion set $M_i$.
\end{itemize}
We labeled a minimum number of contagions for each category by hand for training in the beginning. The number of manually annotated contagions for each category is 100, and the total number of annotated data is 1500. After the initial training process, two classifiers go through the unlabeled contagions to make predictions. If the results from the two classifiers are the same for a contagion, this contagion is added to the labeled set and removed from the unlabeled set. Then a new set for training is obtained, and another iteration starts. In each iteration, there are some contagions moved from the unlabeled set to the labeled set. After enough contagions being labeled, we can derive the following two interactions.

\noindent\textbf{Contagion Category-Category Interaction.}\ \ \
If the set of contagions $ \{m_1, m_2,...,m_k\}$ belongs to category $c_i$, the latent topic distribution $ \varphi_{i,a}(a \in 1,..|t|)$ of category $c_i$  can be obtained through $\frac{\sum_{j=1}^{k}\theta_{j,a}}{k}$. We define $\Lambda_{cate.} \in R^{|c| \times |c|}$ to denote \emph{Contagion Category-Category Interaction}, where $\Lambda_{cate.}(c_i, c_k)$ is the impact of category $c_k$ on $c_i$, that is

\begin{equation}
\small
\Lambda_{categ.}(c_i, c_k) = \sum_a\sum_b\varphi_{i,a}\Lambda_{topic}(t_a,t_b)\varphi_{k,b}
\end{equation}

\noindent\textbf{User Role - Contagion Category Interaction.}\ \ \
Similarly, define $\Omega_{cate.}^{role} \in R ^ {|r| \times |c|}$ to denote \emph{User Role - Contagion Category Interaction}, where $\Omega_{cate.}^{role}(r_i, c_j)$ is the interaction from user role $r_i$ to category $c_j$, that is

\begin{equation}
\small
\Omega_{categ.}^{role}(r_i, c_k) = \sum_b\Omega_{topic}^{role}(r_i,t_b)\varphi_{k,b}
\end{equation}

\section{Sentiment-related Interactions}\label{sec:ldas}

In this section, we first introduce the LDA-S model, and then discuss how to use this model to obtain the sentiment-related interactions.

\subsection{LDA-S Model}

\subsubsection{Motivation and Basic Idea}

We consider how to incorporate sentiments into the IAD framework since a user's forwarding behavior may be affected by the sentiments expressed in the contagions. A natural way is to extract the sentiments from each contagion, and then model their interactions if they are simultaneously exposed to a user. However, extract only sentiments may not be enough as sentiment polarities are usually dependent on topics or domains~\cite{zhao2010jointly}. In other words, the exact same word may express different sentiment polarities for different topics. For example, the opinion word "low" in the phrase "low speed" may have negative orientation in a traffic-related topic. However, if it is in the phrase "low fat" in a food-related topic, the word "low" usually belongs to the positive sentiment polarity. Therefore, it is necessary to incorporate the topic information for sentiment analysis, in order to get topic-specific sentiments, namely topic-sentiments in this work. This can have two benefits. On one hand, extracting the sentiments corresponding to different topics can improve the sentiment classification accuracy. On the other hand, particularly for this work, such method can refine the inferred interactions between topics. Specifically, we use the interactions between topic-sentiments instead of the interactions between topics in the IAD framework, which would improve the accuracy of forwarding prediction.

Inspired by Topic Sentiment Latent Dirichlet Allocation (TSLDA)~\cite{sent5}, we propose LDA-S, an extension of LDA~\cite{LDA} model, to infer sentiment distribution and topic distribution simultaneously for short texts. The difference of LDA-S and TSLDA will be illustrated in Section~\ref{ldadifference}. LDA-S model consists of two steps. The first step aims to obtain the topic distribution of each document (or contagion in our work), and then set the contagion's topic as the one that has the largest probability. The second step gets the sentiment distribution of each document.
The opinion words are usually adjectives or adverbs, whereas the topics words are usually nouns. The words in a document are classified into three categories, the topic words ($c=1$), the sentiment words ($c=2$) and the background words ($c=0$).  We adopt a sentiment word list called~\emph{NTUSD}~\cite{ku2007mining}, which contains 4370 negative words and 4566 positive words. If a word is an adjective but not in the sentiment word list, the sentiment label of this word is set as neutral. If a word is a noun, it is considered as a topic word. Otherwise, it is considered as a background word. In our model, different topics have different opinion word distributions. For each topic, we distinguish opinion word distributions for different sentiment meanings such as positive, negative or neutral. We will show the graphical model and the generation process in the following part.

\subsubsection{Generation Process}
.

\begin{figure}
\centering
\epsfig{file=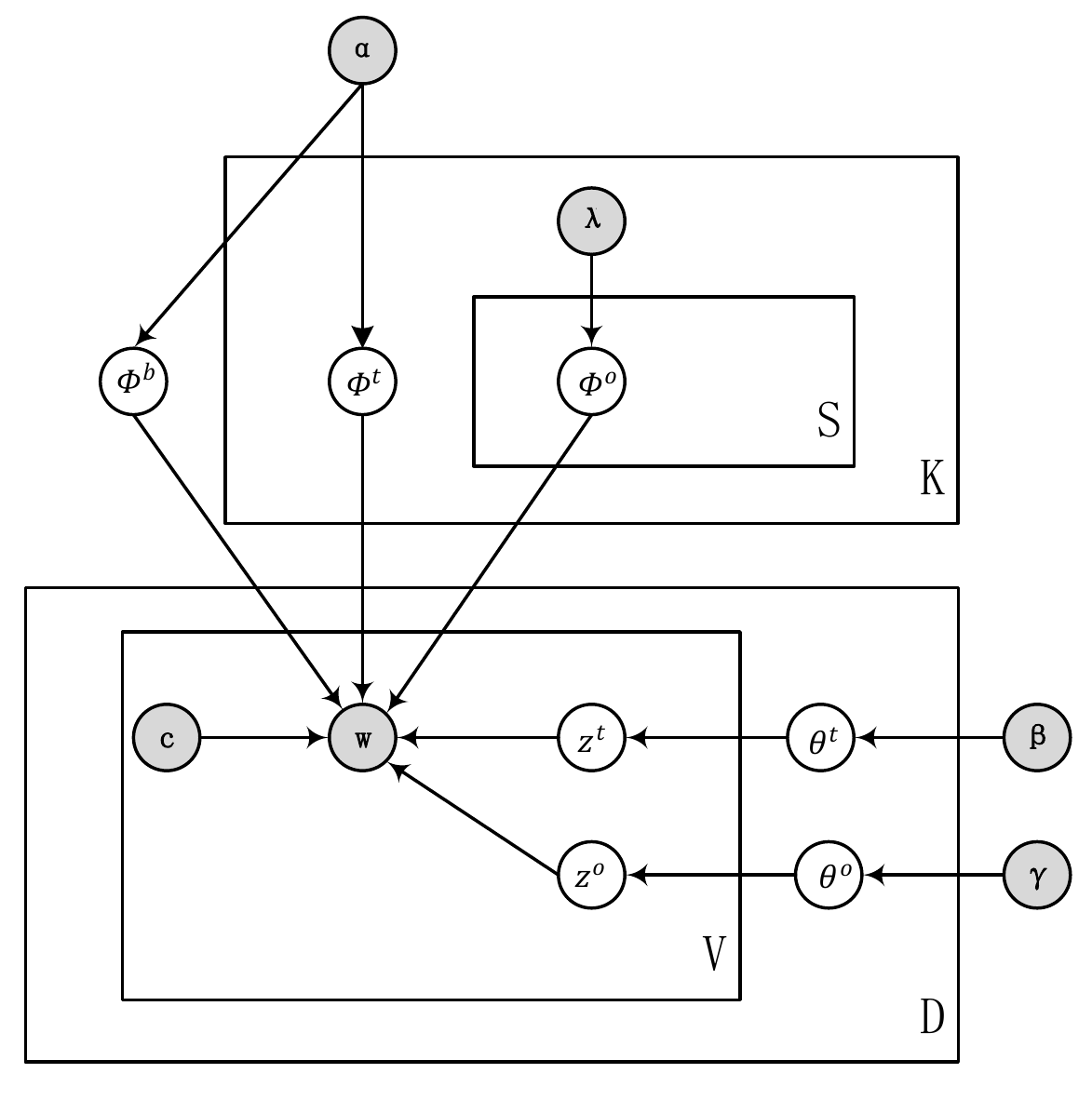, height=2.8in, width=2.8in}
\caption{Graphical Model Representation of LDA-S.}
\label{slda}
\end{figure}

\begin{table}
\small
\centering
\caption{Description of Symbols in LDA-S} \label{sentNotations}
\begin{tabular}{l l} \hline
Symbol &Description\\ \hline
$\alpha,\beta,\gamma,\lambda$ & Dirichlet prior vectors\\
$c$ & Category of words\\
$K$ & \#~of topics\\
$S$ & \#~of sentiments\\
$B$ & \#~of stop words\\
$V$ & \#~of words in corpus\\
$V_0$ & \#~of words with category 0 in corpus\\
$V_1$ & \#~of words with category 1 in corpus\\
$V_2$ & \#~of words with category 2 in corpus\\
$\Phi^b$ & Distribution over background words\\
$\Phi^t$ & Distribution over topic words\\
$\Phi^o$ & Distribution over sentiment words\\
$\theta_d^t$ & Topic distribution for document $d$\\
$\theta_d^o$ & Sentiment distribution for document $d$\\
$\theta_d^{t-o}$ & Topic-Sentiment distribution for document $d$\\
$z^t$ & Topic assignment\\
$z^o$ & Sentiment assignment\\
\hline
\end{tabular}
\end{table}
Figure~\ref{slda} shows the graphical model of LDA-S. Shaded circles indicate observed variables, and clear circles indicate hidden variables. A word is an item from a vocabulary indexed by $\{1,...,V\}$. We represent words using standard basis vectors that have a single component equal to one and all other components equal to zero. Thus, using superscripts to denote components, the $v$th word in the vocabulary is represented by a $V$-vector $w$ such that $w^v=1$ and $w^u=0$ for $u \ne v$. A document $d$ is a sequence of $N$ words denoted by $\textbf{w}=(w_{d,1},w_{d,2},...,w_{d,N})$, where $w_{d,n}$ is the $n$th word in the sequence. A corpus is a collection of $M$ documents denoted by $D = \{\textbf{w}_1,\textbf{w}_2,...,\textbf{w}_M\}$. Let $z^{t}_{d}$ denote the topic assignment of document $d$ and other notations is introduced in Table~\ref{sentNotations}. The generation process of LDA-S is shown as Algorithm 1.

\begin{algorithm}[t]
\small
\caption{LDA-S}  \label{ldas}
\hspace*{0.01in} {\bf Input:} 
documents \\
\begin{algorithmic}[1]
\State Set the prior distributions $\alpha$, $\beta$, $\lambda$ and $\gamma$
\State Set the number of iterations $n$
\For{$iter =1$~to~$n$ }
	\State Choose a distribution of background words 
	$\Phi^{b}\sim Dirichlet(\alpha)$

	\For{each topic $k$} 
　　		\State Choose a distribution of topic words
		$\Phi_{k}^{t}\sim Dirichlet(\alpha)$
		\For{ each sentiment $s$ of topic $k$} 
			\State Choose a distribution of sentiment words
			$\Phi_{k,s}^{o}\sim Dirichlet(\lambda)$
		\EndFor
	\EndFor

	\For{each document $d$ }
		\State Choose a topic distribution $\theta_{d}^{t}\sim Dirichlet(\beta)$
		\State Choose a sentiment distribution $\theta_{d}^{o}\sim Dirichlet(\gamma)$
　　		\For{ each word $w_{d,n}$ }
　　			\State Determine the category of $w_{d,n}$
　　			\If{ the category is 0 }
　　　　			\State choose a word $w_{d,n}\sim Multinomial(\Phi^{b})$
			\EndIf
　　			\If{ the category is 1} 
　　　　			\State choose a topic assignment $z_{d,n}^{t}\sim Multinomial(\theta_{d}^{t})$, 
				choose a word $w_{d,n}\sim Multinomial(\Phi_{z_{d,n}^{t}}^{t})$
			\EndIf
　　			\If{ the category is 2 	}
　　　　			\State choose a sentiment assignment $z_{d,n}^{o}\sim Multinomial(\theta_{d}^{o})$, 
				choose a word $w_{d,n}\sim Multinomial(\Phi_{z_{d}^{t},z_{{d,n}}^{o}}^{o})$
　　			\EndIf
		\EndFor
	\EndFor
\EndFor

\State \Return $\theta_{d}^{t}$,$\theta_{d}^{o}$
\end{algorithmic}
\end{algorithm}

To generate a document, we will generate every word in order. Based on Fig.~\ref{slda} and the above generation process, we can get the generation probability of each word $w_{d,n}$ in document $d$ depending on the category of $w_{d,n}$ as shown in Eq. (\ref{lda1}).

 \begin{equation}
 \small
\left\{
\begin{aligned}
p(w_{d,n})=p(w_{d,n})                                    &      & {c=0}\\
p(w_{d,n},z^{t}_{d,n}|d)=p(w_{d,n}|z^{t}_{d,n})\cdot p(z^{t}_{d,n}|d)        &      & {c=1}\\
p(w_{d,n},z^{o}_{d,n}|d,z^{t}_{d})=p(w_{d,n}|z^{o}_{d,n},z^{t}_{d})\cdot p(z^{o}_{d,n}|d,z^{t}_{d})     &      & {c=2}\\
\end{aligned}
\right.
\label{lda1}
\end{equation}

Then we get the approximate expression of Eq. (\ref{lda1}) in Eq. (\ref{lda2}), Eq. (\ref{lda3}) and Eq. (\ref{lda4}). We will define some notations to explain the equations. Note that we represent $w_{d,n}$ in document $d$ as $w^v$ in the corpus. Let $B_{w^v}$ denote the number of times that the background word $w^v$ appears in all documents. $n_{k,w^v}$ is the number of times that the topic word $w^v$ with the topic $k$ appears in all documents (i.e., corpus) and $\alpha_{k,w^v}$ specifies the value of $\alpha$ when the topic is $k$ and the word is $w^v$. Let $n_{z^{t}_{d},s,w^v}$ be the number of times that the sentiment word $w^v$ with the sentiment $s$ under the topic $z^{t}_{d}$ appears in all documents and $\lambda_{z^{t}_{d},s,w^v}$ specifies the value of $\lambda$ when the topic is $z^{t}_{d}$, the sentiment is $s$ and the word is $w^v$. $n_{d,k}$ denotes the number of times that the document $d$ is endowed with topic $k$ while $\beta_{d,k}$ specifies the value of $\beta$ when the document is $d$ and the topic is $k$. $n_{d,s}$ is the number of times that the document $d$ is endowed with sentiment $s$ while $\gamma_{d,s}$ specifies the value of $\gamma$ when the document is $d$ and the sentiment is $s$. When $c=0$, we only consider the background words. Specifically, we use $\alpha_{w^v}$ to denote the value of $\alpha$ when the word $w^v$ is a background word, and the background word has no topic assignment in this work. Then the generation probability of $w^v$ is

\begin{equation}
\small
p(w^v) \propto \frac{B_{w^v}+\alpha_{w^v}}{\sum_{v=1}^{V_0}(B_{w^v}+\alpha_{w^v})}
\label{lda2}
\end{equation}

When $c=1$, we only consider the topic words. The generation probability of $w^v$ is

\begin{equation}
\small
\begin{split}
p(w^v,k|d)=p(w^v|k)\cdot p(k|d) \\
\propto \frac{n_{k,w^v}+\alpha_{k,w^v}}{\sum_{v=1}^{V_1}(n_{k,w^v}+\alpha_{k,v})}   \\
\cdot \frac{n_{d,k}+\beta_{d,k}}{\sum_{k=1}^{K}(n_{d,k}+\beta_{d,k})}  \\
\end{split}
\label{lda3}
\end{equation}

When $c=2$, we only consider the sentiment words. When the topic of document $d$ is $z^{t}_{d}$, the generation probability of $w^v$ is

\begin{equation}
\small
\begin{split}
p(w^v,s|d,z^{t}_{d})=p(w^v|s, z^{t}_{d})\cdot p(s|d,z^{t}_{d})    \\
 \propto \frac{n_{z^{t}_{d},s,w^v}+\lambda_{z^{t}_{d},s}}{\sum_{v=1}^{V_2}(n_{z^{t}_{d},s,w^v}+\lambda_{z^{t}_{d},s,w^v})} \\
 \cdot \frac{n_{d,s}+\gamma_{d,s}}{\sum_{s=1}^{S}(n_{d,s}+\gamma_{d,s})} \\
\label{lda4}
\end{split}
\end{equation}

Then Eq. (\ref{lda3}) and Gibbs Sampling are implemented for inference in LDA-S in which we only consider words with category 1. It will sequentially sample the hidden variables $z_{d,n}^t$ (the topic assignment of  the $n$th word in document $d$) given a set of observed variables $\textbf{w}$ and a set of hidden variables ${\textbf{z}_{-(d,n)}^t}$. A bold-front variable denotes the list of the variables. For example, $\textbf{w}$ denotes all the words in all documents. ${\textbf{z}_{-(d,n)}^t}$ denotes all topic assignment variables $z^t$ except $z_{d,n}^t$. $-(w^v)$ stands for all the words except $w^v$. $n_{d,k,-w^v}$ denotes the number of times that words are endowed with the topic $k$ except the word $w^v$. ${\textbf{w}_{-w^v}}$ denotes all the words in all documents except the word $w^v$. The equation of topic sampling is shown in Eq. (\ref{lda5}).

\begin{equation}
\small
\begin{split}
p(z_{d,n}^{t}=k|\textbf{z}_{-(d,n)}^{t},{\textbf{w}}) \\
\propto p(z_{d,n}^{t}=k, w_{d,n}=w^v|\textbf{z}_{-(d,n)}^{t},{\textbf{w}_{-w^v}}) \\
\propto \frac{n_{d,k,-n}+\beta_{d,k}}{\sum_{k=1}^{K}(n_{d,k,-n}+\beta_{d,k})} \\
\cdot \frac{n_{k,-w^v}+\alpha_{k,w^v}}{\sum_{v=1}^{V_1}(n_{k,-w^v}+\alpha_{k,w^v})}
\end{split}
\label{lda5}
\end{equation}

After we get the topic distribution of document $d$, the topic that has the largest probability will be set as document $d$'s topic. For example, if $\theta_d^t=(0.1, 0.2, 0.3, 0.4)$, we will choose the last one as document $d$'s topic. If there are more than one topics with the same probability, we choose one of them randomly. Then Gibbs Sampling will be adopted again together with  Eq. (\ref{lda4}), in which we only consider words with category 2. We use the observed variables $\textbf{w}$, $z^{t}_{d}$ as well as the hidden variable ${\textbf{z}_{-(d,n)}^o}$ to sample the hidden variable $z_{d,n}^o$  which denotes the sentiment assignment of $n$th word in document $d$. The equation of sentiment sampling is shown in Eq. (\ref{lda6}).

\begin{equation}
\small
\begin{split}
p(z_{d,n}^{o}=s|\textbf{z}_{-(d,n)}^{o},{\textbf{w}},z^{t}_{d}) \\
\propto p(z_{d,n}^{o}=s, w_n=w^v|\textbf{z}_{-(d,n)}^{o},{\textbf{w}_{-w^v}},z^{t}_{d}) \\
\propto \frac{n_{d,s,-n}+\gamma_{d,s}}{\sum_{s=1}^{S}(n_{d,s,-n}+\gamma_{d,s})} \\
 \cdot\frac{n_{z^{t}_{d},s,-w^v}+\lambda_{z^{t}_{d},s,w^v}}{\sum_{v=1}^{V_2}(n_{z^{t}_{d},s,-w^v}+\lambda_{z^{t}_{d},s,w^v})}
\end{split}
\label{lda6}
\end{equation}

After that, we can approximate the multinomial parameter sets with the samples from Eq. (\ref{lda5}) and Eq. (\ref{lda6}). The distributions of topics and sentiments in document $d$ are shown in Eq. (\ref{lda7a}) and Eq. (\ref{lda7b}) respectively,
\begin{equation}
\small
\begin{split}
\theta_{d,a}^t=\frac{n_{d,a}+\beta_{d,a}}{\sum_{k=1}^{K}(n_{d,k}+\beta_{d,k})}
\end{split}
\label{lda7a}
\end{equation}

\begin{equation}
\small
\begin{split}
\theta_{d,b}^o=\frac{n_{d,b}+\gamma_{d,b}}{\sum_{s=1}^{S}(n_{d,s}+\gamma_{d,s})}
\end{split}
\label{lda7b}
\end{equation}
where $a$ denotes one of the topics, and $b$ denotes one of the sentiments. Then the joint distribution of topic $a$ and sentiment $b$ is obtained in Eq. (\ref{lda8}).

\begin{equation}
\small
\theta_{d,ab}^{t-o}=\theta_{d,a}^t \cdot \theta_{d,b}^o
\label{lda8}
\end{equation}

The background word distribution is shown in Eq. (\ref{lda9}) in which we only consider the background words, the topic word distribution of topic $k$ is shown in Eq. (\ref{lda10}) in which we only consider the topic words and the sentiment word distribution of sentiment $s$ under the topic $k$ is shown in Eq. (\ref{lda11}) in which we only consider the sentiment words, where $r$ denotes one of the words in the corpus.

\begin{equation}
\small
\Phi_r^b=\frac{B_r+\alpha_r}{\sum_{v=1}^{V_0}(B_{v}+\alpha_{v})} \\
\label{lda9}
\end{equation}

\begin{equation}
\small
\Phi_{k,r}^t=\frac{n_{k,r}+\alpha_{k,r}}{\sum_{v=1}^{V_1}(n_{k,v}+\alpha_{k,v})}\\
\label{lda10}
\end{equation}

\begin{equation}
\small
\Phi_{r,k,s}^o=\frac{n_{k,s,r}+\lambda_{k,s,r}}{\sum_{v=1}^{V_2}(n_{k,s,v}+\lambda_{k,s,v})}\\
\label{lda11}
\end{equation}

\subsubsection{With Prior Information}
\label{prior}
For the hyperparameters of this LDA-S, namely, $\alpha$, $\beta$, $\lambda$ and $\gamma$, we set $\alpha$=0.1 and $\beta$=0.01 as TSLDA~\cite{sent5}. But for prior $\lambda$ and $\gamma$, we adopt a sentiment word list to obtain them. The intuition is that incorporating prior information or subjective lexicon (i.e., words bearing positive and negative polarity) are able to improve the sentiment detection accuracy~\cite{sent4}. The prior parameters of Dirichlet distribution can be regarded as the ``pseudo-counts" from ``pseudo-data". To get the priors, we match the documents to the sentiment word list. Then the number of the words with a specific sentiment $s$ appearing in a document $d$ can be obtained, which can be set as the value of  $\gamma_{d,s}$. Similarly, the value of $\lambda_{k,v,s}$ can be set as the count of the word $v$ with sentiment $s$ under topic $k$. With the prior information, our experimental results show a significant improvement in the sentiment classification accuracy, but the results are omitted due to the limit of space.


\subsubsection{Comparision with TSLDA}
\label{ldadifference}

TSLDA is designed for long documents which consist of a set of sentences, and is not suitable to short texts with only one or two sentences. Specifically, the limit in length makes it difficult to achieve good learning results. Compared to TSLDA, LDA-S adopts different assumptions and different sampling methods. Moreover, LDA-S incorporates prior information at the beginning of the generation process, which is expected to be more suitable to short texts. The differences are listed as follows.

\begin{itemize}
\item \textbf{Sampling Method.} With TSLDA, each sentence is supposed to express only one topic and one opinion on that topic, and we need to sample topics and sentiments for each sentence in the generation process. But in LDA-S, each contagion (which usually consists of one or two sentences) bears a topic distribution and a sentiment distribution, and we don't involve sentences in the generation process.\\

\item \textbf{Prior Information.} In the initialization step of TSLDA, each document and word are assigned with topics and sentiments randomly, limiting its performance.  In LDA-S, we use a paradigm word list consists of a set of positive and negative words, and compare each word token in the contagions against the words in the sentiment word list in the initialization of the generation process.

\end{itemize}

\subsection{Interactions with Sentiment}

After the topic-sentiment distribution (as shown in Eq.~(\ref{lda8})) is obtained, the \emph{Contagion (Topic-Sentiment) - (Topic-Sentiment) Interaction} and \emph{User Role - Contagion (Topic-Sentiment) Interaction} will be introduced in our framework. Now, $\pi(x_i)$ (i.e., Eq. (\ref{eqn:goal})) can be updated by $\Omega_{t-s}^{role}$, $\Delta_{role}$ and $\Lambda_{t-s}$.

Specifically, each contagion is assumed to have a distribution not only on topics but also on sentiments, and $ts$ denotes the set of tuples that contain a latent topic and a kind of sentiment. LDA-S is used to extract the joint distribution of latent topic and sentiment on each contagion. Therefore, the value of $|ts|$ is actually the number of topics (20 in our setting) times the number of sentiment polarities (3 in our setting). Then, instead of modeling $\Lambda_{topic} \in R^{ |t| \times |t|}$, we would model a matrix $\Lambda_{t-s} \in R^{ |ts| \times |ts|}$, which denotes the \emph{Contagion (Topic-Sentiment) - (Topic-Sentiment) Interaction}. We define $ \theta_{i,ax}$ as the probability of contagion $m_i$ belonging to topic $t_a$ and sentiment $s_x$, and therefore $\sum_{ax}\theta_{i,ax}=1$. Let $\Lambda_{t-s}(t_{ax},t_{by})$ denote the impact of topic-sentiment $t_{by}$ has on topic-sentiment $t_{ax}$. Then $ \Lambda(m_i, m_k)$  in Eq. (\ref{eqn:c-inter}) can be updated by

\begin{equation}
\small
\Lambda(m_i, m_k) = \sum_{ax}\sum_{by}\theta_{i,ax}\Lambda_{t-s}(t_{ax},t_{by})\theta_{k,by}
\label{eqn:topic-inter}
\end{equation}


Instead of learning $\Omega$, we build a matrix $\Omega_{t-s}^{role} \in R^{ |r| \times |ts|}$ to denote the \emph{User Role - Contagion (Topic-Sentiment) Interactions}. Then $\Omega(u_a,m_i)$ in Eq. (\ref{eqn:u-c-inter}), Eq. (\ref{eqn:u-inter}) and Eq. (\ref{eqn:c-inter}) can be updated by

\begin{equation}
\small
\Omega(u_a,m_i) = \sum_j\sum_{by}\vartheta_{a,j}\Omega_{t-s}^{role}(r_j,t_{by})\theta_{i,{by}}
\label{eqn:role-topic-inter}
\end{equation}

Now, $\pi(x_i)$ can be updated by $\Omega_{t-s}^{role}$, $\Delta_{role}$ and $\Lambda_{t-s}$,  and the log-likelihood function is

\begin{equation}
\small
\begin{split}
&L(\Omega_{t-s}^{role}, \Delta_{role}, \Lambda_{t-s}) \\
= & \sum_{i=1}^n(y_i log \pi(x_i)+(1-y_i)log(1-\pi(x_i)))
\label{eqn:sentlikelihood}
\end{split}
\end{equation}

Our goal is to estimate the parameters in $\Omega_{t-s}^{role}$, $\Delta_{role}$ and $\Lambda_{t-s}$ to maximize the log-likelihood function. Stochastic gradient ascent is adopted again to fit the model. Then $\Omega_{s}^{role}(r_j,t_{y})$ and  $\Lambda_{s}(t_x, t_y)$ can be obtained by

\begin{equation}
\small
\Omega_{s}^{role}(r_j,t_y) = \frac{1}{K}\sum_b\Omega_{t-s}^{role}(r_j,t_{by})
\end{equation}


\begin{equation}
\small
\Lambda_{s}(t_x,t_y) = \frac{1}{K}\sum_a\frac{1}{K}\sum_b\Lambda_{t-s}(t_{ax},t_{by})
\end{equation}



%
%
%
%

\section{Evaluation}

In this section, we conduct experiments based on a public Weibo dataset to evaluate IAD framework, and then discuss various qualitative insights.

\subsection{Experimental Settings}

\subsubsection{Dataset}
Weibo is a Twitter-like social network which provides microblogging service. The Weibo dataset~\cite{weibodataset} provides a list of users who have forwarded contagions, as well as the forwarding timestamp. Users' friendship links are also recorded. Because of the crawling strategy, the distribution of retweet counts in different months is highly imbalanced. Thus, we select the diffusion data from July 2012 to December 2012, in which the retweet count per month is large enough and the distribution is more balanced. Consequently, we get 19,388,727 retweets on 140,400 popular microblogs. We delete the inactive users without any retweets in this period and obtain 1,077,021 distinct users for the experiment.

Then we do statistical analysis to extract interacting scenarios from the dataset as illustrated in Sec.~\ref{sec:model}. Please note that we remove the interacting scenarios where all microblogs are neutral, and obtain a set of scenarios that have contagions with positive or negative sentiments. As LDA-S is proposed to study how the sentiments can affect the diffusion process, thus the contagions that we focus on should be those whose sentiments are evident. Based on this intuition, we refine the dataset and only retain the contagions whose positive portion or negative portion in the sentiment distribution is above 0.7, and thus obtain a new dataset consisting of contagions with greater sentiment intensities. Thus, the obtained instances after filtering is actually a subset of the instances used in the earlier version of this work~\cite{suunderstanding}. If the examined contagion is adopted, the interacting scenario is a positive instance, otherwise it is a negative instance. We observe that the positive and negative instances are highly unbalanced in the dataset, so we sample a balanced dataset with equal numbers of positive and negative instances. In total, we get 64,456 microblogs and 14,537,835 interacting scenarios when $K=1$, while 75,945 microblogs and 17,462,853 interacting scenarios when $K=2$. We use 5-fold cross-validation, and set the number of latent topics as $|t|=20$.
\\

\subsubsection{Baselines}
We compare the performance of our proposed model with several existing models. These methods are:

\begin{itemize}
\item \textbf{IP Model.} Infection Probability Model is based on the Independent Cascade Model~\cite{Maximizing,Talk}, which assigns the infection probability of a contagion to be simply the prior infection probability. IP model doesn't consider any interactions among users and contagions.   \\

\item \textbf{UI Model.} User Interaction (UI) Model is actually one component of our IAD framework. UI model only considers the user-user interactions, more specifically, the user role-role interactions, and ignores the other interactions.

\item \textbf{IMM Model~\cite{Clash}.} IMM incorporates the interactions among contagions into its model. IMM models the interactions between clusters (i.e., latent topics) of contagions. At the same time, it learns to which cluster each contagion belongs to. However, the clusters (i.e., latent topics) cannot be interpreted and thus prevent us getting deeper insights.

\item \textbf{IAD w/o S.} IAD w/o S refers to IAD framework without sentiment, which is also the proposal in our previous work~\cite{suunderstanding}.

\item \textbf{IAD w/ TSLDA.} In our proposed IAD framework, we replace LDA-S with TSLDA algorithm~\cite{sent5}, which intends to compare the performance difference between LDA-S and TSLDA.

\end{itemize}

To make a fair comparison, we use the same set of instances and the same setting of parameters. In our proposal and the baselines, we set the predicting result to 0 if the predicting infection probability is less than 0.5, otherwise we set the predicting result to 1. Our model and the baselines are evaluated in terms of Precision, Recall, F1-score, as well as Accuracy. All experiments are performed on a dual-core Xeon E5-2620 v3 processor. The code of our proposal has been made publicly available via~\url{https://www.dropbox.com/s/lp296h2omhhfw6e/IAD.zip?dl=0}.

\subsection{Performance Results}

\subsubsection{Effectiveness}

\begin{table}[!t]
\begin{center}
\small
\caption{Performance of IAD Methods Compared to Baselines When $K=1$ }
\label{Effectiveness1}
\centering
\begin{adjustbox}{max width=0.98\columnwidth}
\begin{tabular}{ccccc} \hline
Model Name & Precision & Recall & Accuracy & F1-score \\ \hline
IP &0.7042 &0.6338 &0.6789 &0.6671\\
UI &0.7534&0.6134 &0.6923 &0.6762\\
IMM &0.8005 &0.6376 &0.7234 &0.7098\\
IAD w/o S ($|t|$ = 20) &0.8112 &0.6505 &0.7332 &0.7220\\
IAD w/ TSLDA ($|t|$ = 20) &0.8156 &0.6579 &0.7391 &0.7293\\
IAD w/ LDA-S ($|t|$ = 20) &0.8243 &0.6608 &0.7439 &0.7365\\ \hline
\end{tabular}
\end{adjustbox}
\end{center}
\end{table}

\begin{table}
\begin{center}
\small
\caption{Performance of IAD Methods Compared to Baselines When $K=2$  }
\label{Effectiveness2}
\begin{adjustbox}{max width=0.98\columnwidth}
\begin{tabular}{ccccc} \hline
Model Name & Precision & Recall & Accuracy  & F1-score \\ \hline
IP &0.7124 &0.6031 &0.6799 &0.6532\\
UI &0.7221 &0.6089 &0.6846 &0.6607\\
IMM &0.7345 &0.6275 &0.6954 &0.6768\\


IAD w/o S ($|t|$ = 20) &0.7616 &0.6559 &0.7178 &0.7048\\
IAD w/ TSLDA ($|t|$ = 20) &0.7757 &0.6601	&0.7267 &0.7132\\
IAD w/ LDA-S ($|t|$ = 20) &0.7775 &0.6664	&0.7298 &0.7177\\ \hline
\end{tabular}
\end{adjustbox}
\end{center}
\end{table}

Table~\ref{Effectiveness1} and Table~\ref{Effectiveness2} show the performance of our proposal and the baselines when $K=1$ and $K=2$ respectively. IAD w/ LDA-S denotes the IAD framework with our LDA-S method for extracting sentiments. It can be observed that UI and IMM models perform better than IP model, indicating that interactions among users and among contagions are vital to user's forwarding decisions. Moreover, since IMM outperforms UI, the interaction among contagions plays more important roles than the interaction among users. It can be observed that our proposed IAD methods (no matter with or without sentiments) almost consistently outperform IP, UI and IMM, indicating that solely involving the interactions between contagions (e.g., IMM model) or between user interactions (e.g., UI) is not sufficient. Besides, we need to involve both of the aforementioned interactions as well as the interactions between users and contagions to get a more accurate prediction. Then we examine the effectiveness of the sentiment factors in the forwarding decision process, and observe that IAD w/ LDA-S and IAD w/ TSLDA both perform better than IAD w/o S in terms of F1-score when $K=1$ and $K=2$. Thus, it indicates that involving sentiments into the the framework is beneficial to the information diffusion prediction. When comparing the performance of IAD w/ LDA-S and IAD w/ TSLDA, we can find that IAD w/ LDA-S outperforms IAD w/ TSLDA all the time in terms of both F1-score and Accuracy, indicating that our proposed LDA-S can better capture the effects of the sentiments on the information diffusion than TSLDA. We also conduct experiments with a larger $K$, and the results are not better than those when $K=1$ and $K=2$. The possible reason is that individual users have limited attention, and can only jointly consider a very limited number of contagions.

\begin{table}
\begin{center}
\small
\caption{Performance of IAD Methods with Varying Sentiment Intensities When $K=1$}
\label{Intensities1}
\begin{adjustbox}{max width=0.98\columnwidth}
\begin{tabular}{ccccc} \hline
Model Name & Precision & Recall & Accuracy  & F1-score \\ \hline
IAD w/o S ($|t|$=10, $\tau$=0.6) &0.7759 &0.6589 &0.7223 &0.7126\\
IAD w/ LDA-S ($|t|$=10, $\tau$=0.6) &0.8156 &0.6411 &0.7250 &0.7179\\
IAD w/o S ($|t|$=10, $\tau$=0.7) &0.7734 &0.6657 &0.7243 &0.7155\\
IAD w/ LDA-S ($|t|$=10, $\tau$=0.7) &0.8187 &0.6442 &0.7343 &0.7210\\
IAD w/o S ($|t|$=10, $\tau$=0.8) &0.7707 &0.6512	&0.7198 &0.7059\\
IAD w/ LDA-S ($|t|$=10, $\tau$=0.8) &0.8205 &0.6483	&0.7379 &0.7243\\ 
IAD w/o S ($|t|$=20, $\tau$=0.6) &0.8167 &0.6564 &0.7386 &0.7278\\
IAD w/ LDA-S ($|t|$=20, $\tau$=0.6) &0.8208 &0.6592 &0.7413 &0.7312\\
IAD w/o S ($|t|$=20, $\tau$=0.7) &0.8112 &0.6505 &0.7332 &0.7220\\
IAD w/ LDA-S ($|t|$=20, $\tau$=0.7) &0.8243 &0.6608 &0.7439 &0.7365\\
IAD w/o S ($|t|$=20, $\tau$=0.8) &0.8087 &0.6495	&0.7348 &0.7204\\
IAD w/ LDA-S ($|t|$=20, $\tau$=0.8) &0.8259 &0.6657	&0.7475 &0.7372\\ \hline
\end{tabular}
\end{adjustbox}
\end{center}
\end{table}

\begin{table}
\begin{center}
\small
\caption{Performance of IAD Methods with Varying Sentiment Intensities When $K=2$}
\label{Intensities2}
\begin{adjustbox}{max width=0.98\columnwidth}
\begin{tabular}{ccccc} \hline
Model Name & Precision & Recall & Accuracy  & F1-score \\ \hline
IAD w/o S ($|t|$=10, $\tau$=0.6) &0.7459 &0.6502 &0.7137 &0.6948\\
IAD w/ LDA-S ($|t|$=10, $\tau$=0.6) &0.7608 &0.6538 &0.7186 &0.7033\\
IAD w/o S ($|t|$=10, $\tau$=0.7) &0.7423 &0.6478 &0.7103 &0.6918\\
IAD w/ LDA-S ($|t|$=10, $\tau$=0.7) &0.7621 &0.6576 &0.7212 &0.7060\\
IAD w/o S ($|t|$=10, $\tau$=0.8) &0.7395 &0.6443	&0.7076 &0.6886\\
IAD w/ LDA-S ($|t|$=10, $\tau$=0.8) &0.7658 &0.6595	&0.7237 &0.7087\\ 
IAD w/o S ($|t|$=20, $\tau$=0.6) &0.7651 &0.6590 &0.7221 &0.7081\\
IAD w/ LDA-S ($|t|$=20, $\tau$=0.6) &0.7746 &0.6632 &0.7263 &0.7146\\
IAD w/o S ($|t|$=20, $\tau$=0.7) &0.7616 &0.6559 &0.7178 &0.7048\\
IAD w/ LDA-S ($|t|$=20, $\tau$=0.7) &0.7775 &0.6664 &0.7298 &0.7177\\
IAD w/o S ($|t|$=20, $\tau$=0.8) &0.7579 &0.6510	&0.7139 &0.7004\\
IAD w/ LDA-S ($|t|$=20, $\tau$=0.8) &0.7789 &0.6678	&0.7306 &0.7191\\ \hline

\end{tabular}
\end{adjustbox}
\end{center}
\end{table}

Moreover, to demonstrate how the sentiment intensity affects the prediction performance, we construct three datasets whose contagions are with different sentiment intensities. Specifically, the positive or negative portions in the sentiment distribution in the three datasets are above $\tau$=0.6, $\tau$=0.7 and $\tau$=0.8, and we denote them as the dataset (a), (b) and (c) respectively for simplicity. It is obvious that the sentiment intensities of contagions in (c) are generally greater than that in (a) and (b). We perform the experiments on the three datasets, and can observe that our proposal performs better on the dataset whose sentiment intensities is greater. For example, in terms of accuracy, when $|t|$=10, the performance gaps between IAD with LDA-S and without LDA-S are 0.27\%, 1\% and 1.81\% using dataset (a) and (b) and (c) respectively; when $|t|$=20, the corresponding performance gaps are 0.27\%, 1.07\% and 1.27\%.  It is intuitive as LDA-S is proposed to study how the sentiments can affect the diffusion process, and thus applying LDA-S on the contagions whose sentiments are evident is supposed to be more beneficial to the prediction than the contagions whose sentiments are mostly neutral. The detailed results are shown in Table~\ref{Intensities1} and Table~\ref{Intensities2}. These results indicate that sentiment is an important factor for information diffusion. Please note that in addition to improving the predictive capability, taking sentiments into consideration can also provide a new perspective to understand the information diffusion process, and the analysis is described in Sec.~\ref{6.3}.

\subsubsection{Efficiency}

\begin{figure}
  \centering
    \epsfig{file=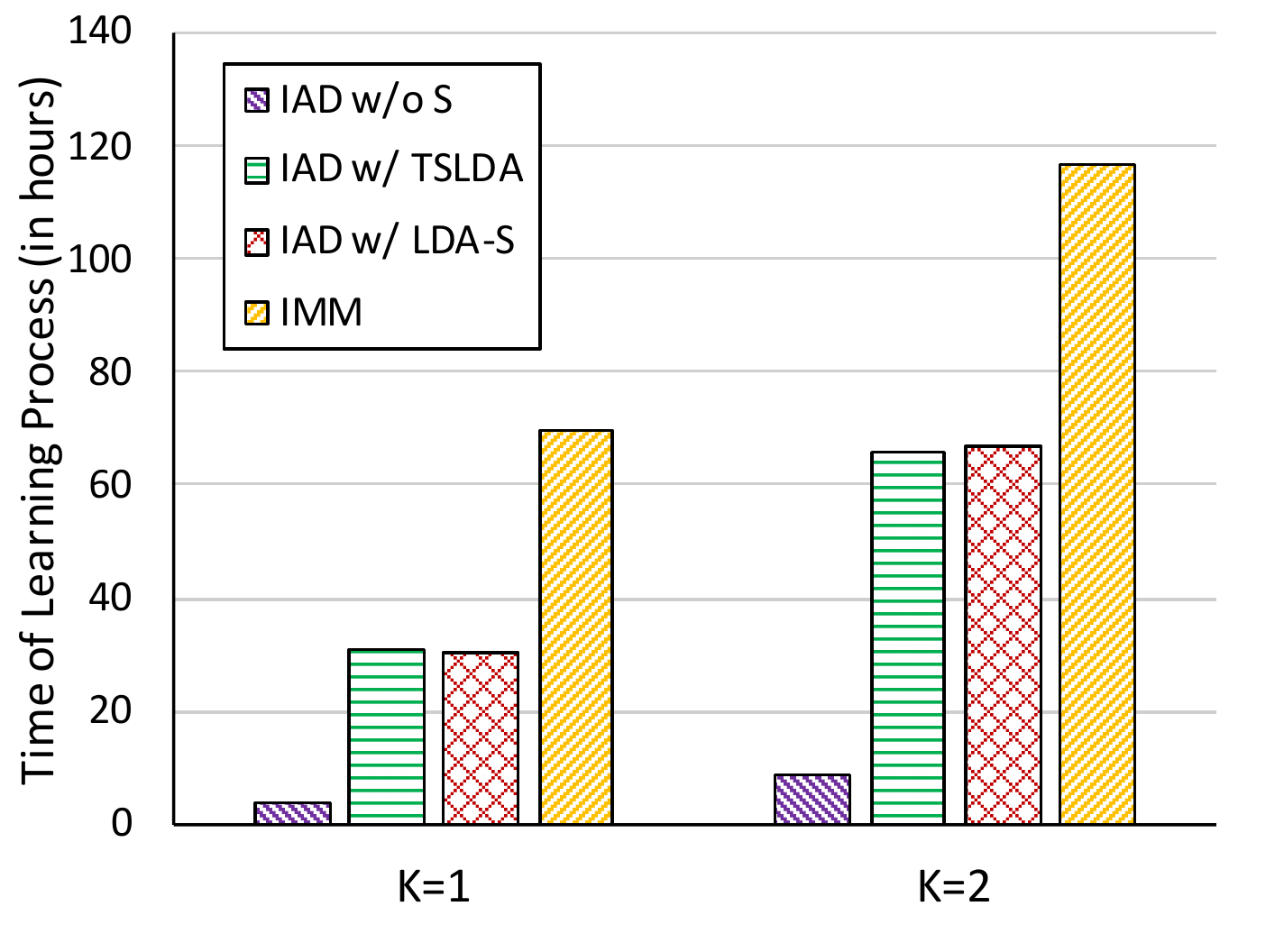, height=1.4in, width=2.1in}

  \caption{Efficiency of IAD Compared to IMM}
  \label{time}
\end{figure}

Taking the model complexity into consideration, IAD w/o S is much more efficient than IMM. The number of parameters to learn in IAD w/o S is 469 (when $|r| = 3$, and $|t| = 20$), whereas the corresponding number in IMM model is 2,070,520 and 2,141,740 respectively when $K=1$ and $K=2$ (obtained by $T \times T \times K + W \times T$, where $T = 20$, $W = 103,506$ and $107,047$ when $K=1$ and $K=2$). For IAD w/ LDA-S, the number of parameters to infer is 3789. Fig.~\ref{time} compares the time cost in the learning process, and the results confirm the efficiency of the IAD models, especially the IAD w/o S model. In particular, IAD w/o S is faster than IMM by an order of magnitude. IAD w/ LDA-S and IAD w/ TSLDA take almost the same time in learning, and they are both slower than IAD w/o S but much faster than IMM. When $K=2$, similar trends can be observed.
%


\subsection{Analysis of Interactions}\label{6.3}

\begin{figure*}[!t]
  \centering
  \subfloat[User Role-Role Interactions]{
  \includegraphics[height=2.0in, width=2.3in]{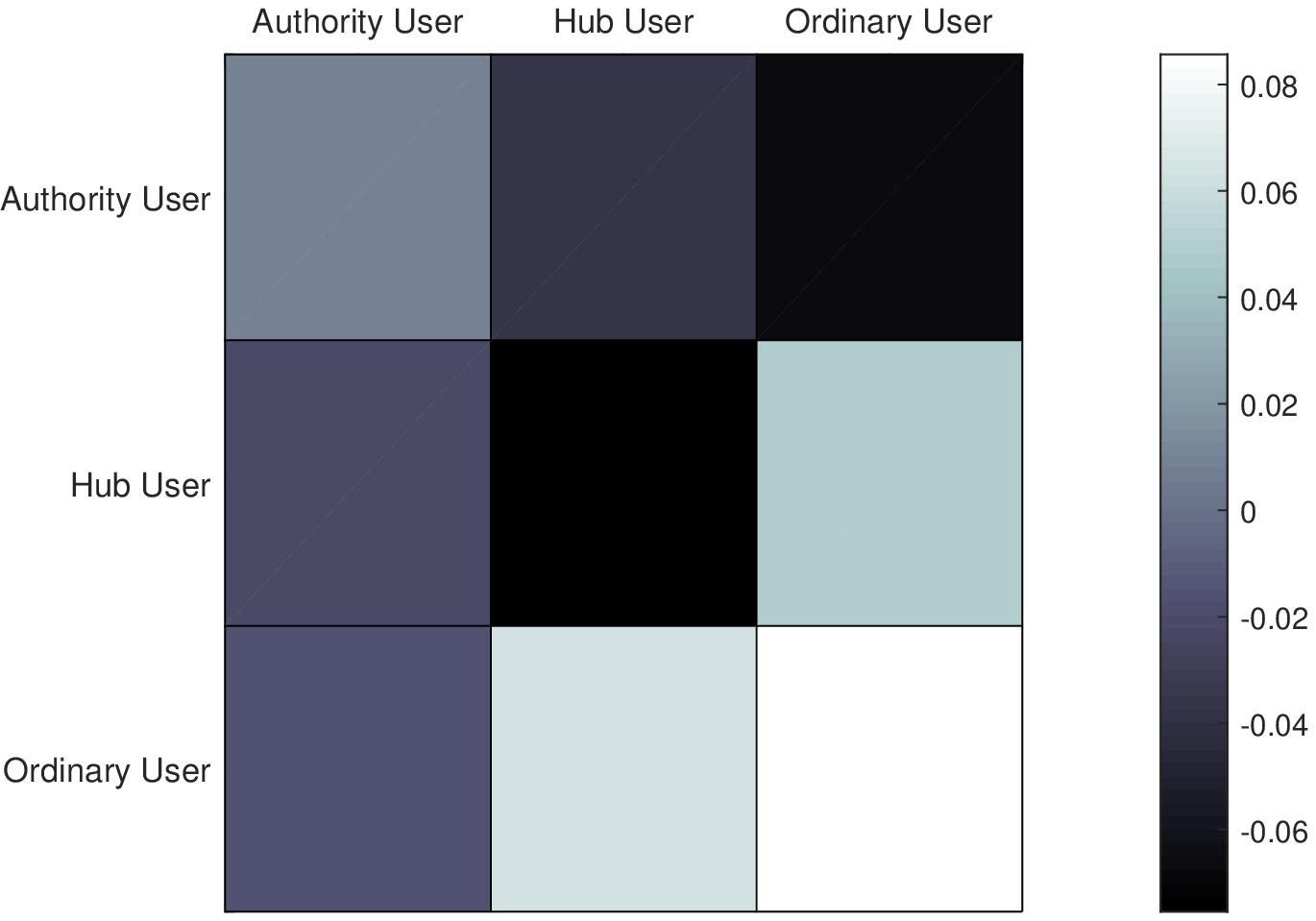}}
 \subfloat[User Role - Contagion Sentiment Interactions]{
  \includegraphics[height=2.0in, width=2.3in]{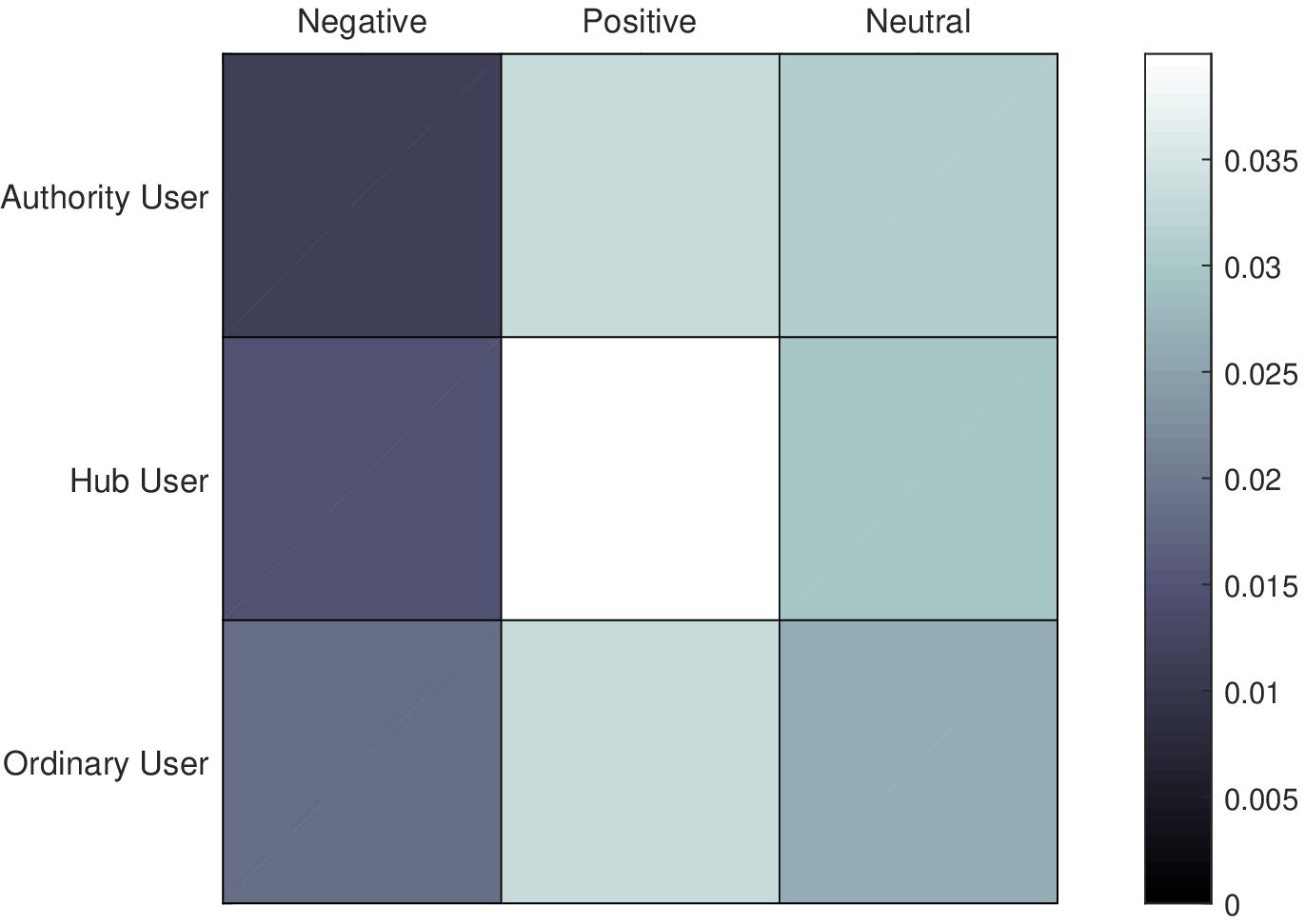}}
  \subfloat[Contagion Sentiment-Sentiment Interactions]{
   \includegraphics[height=2.0in, width=2.3in]{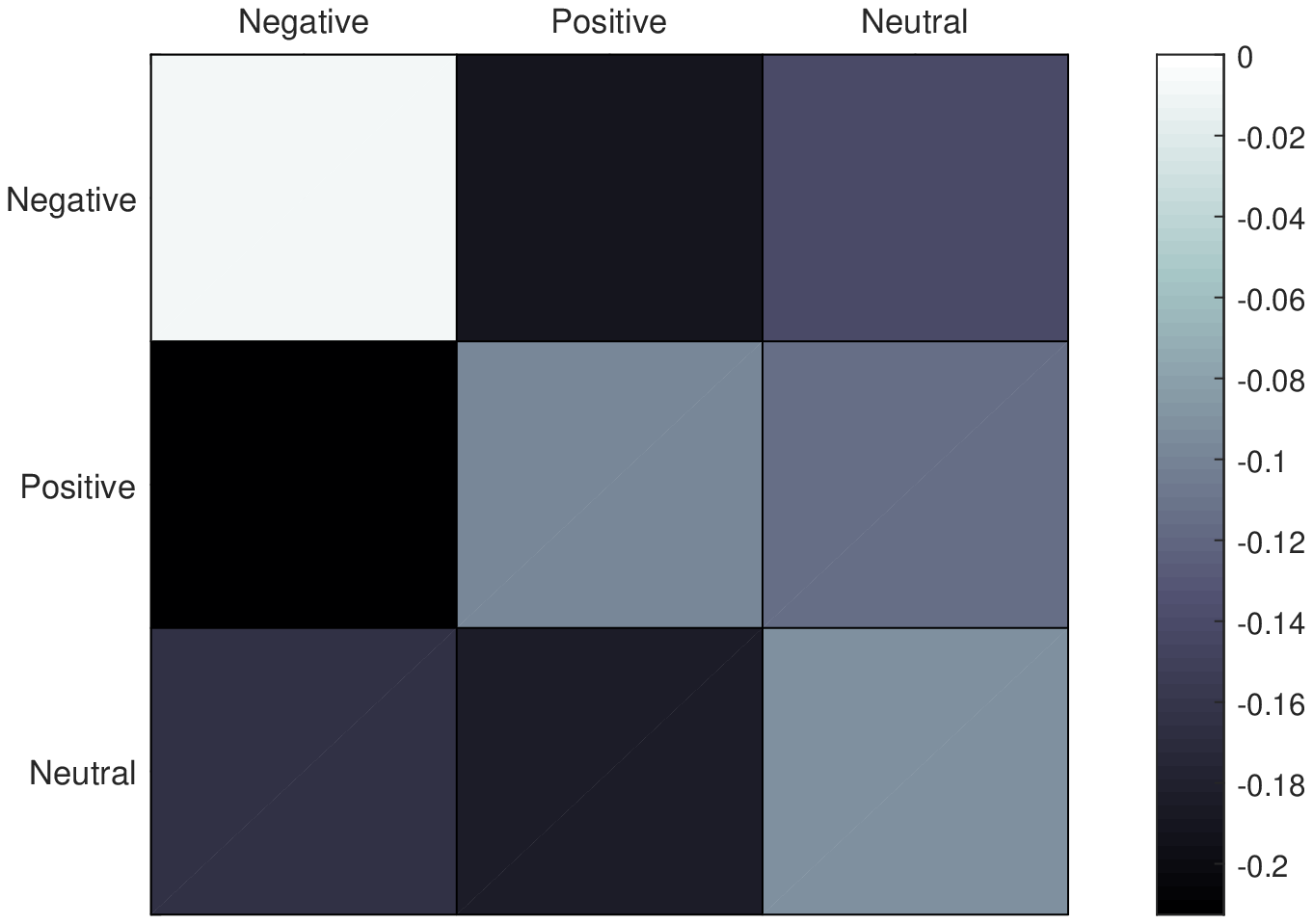}}

      \subfloat[User Role - Contagion Category Interactions ]{
       \includegraphics[height=2.2in, width=3.0in]{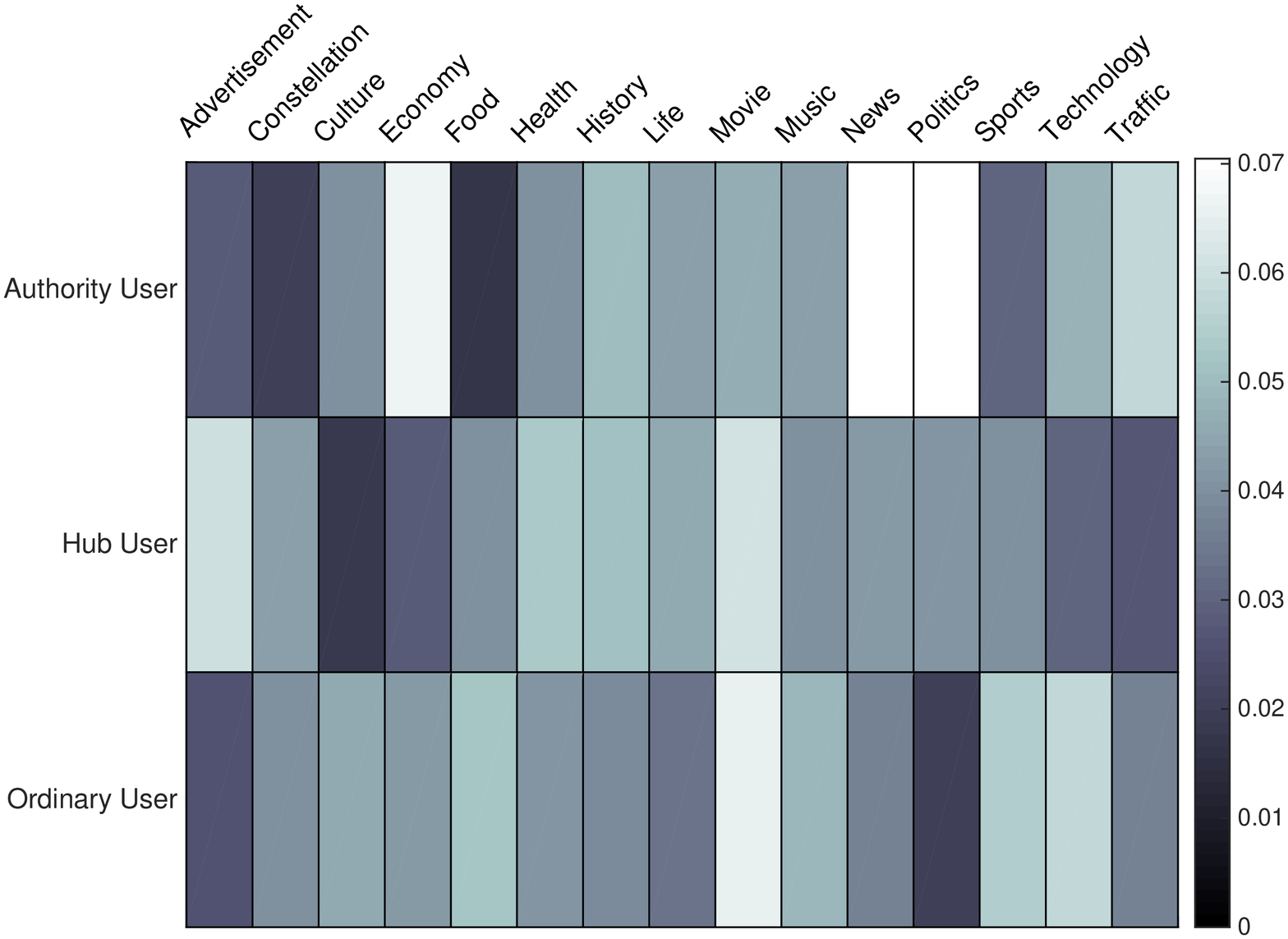}}
  \subfloat[Contagion Category-Category Interactions]{
   \includegraphics[height=2.2in, width=4.0in]{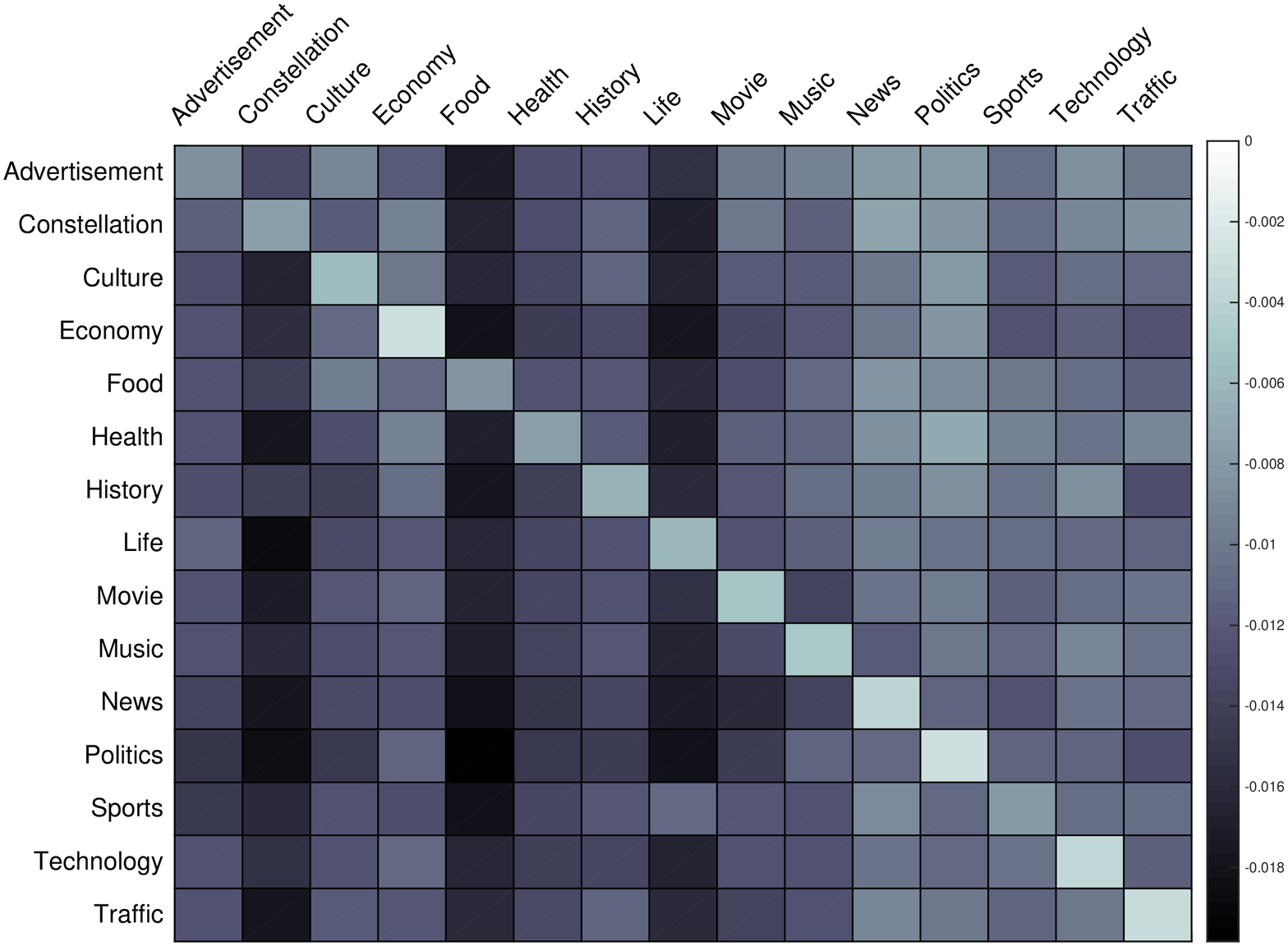}}

  \caption{The inferred interactions when $K=2$ and $|t|=20$. (a) User Role-Role Interaction $\Delta_{role}(r_i, r_j)$, with $r_i$ as the ordinate and $r_j$ as the abscissa, denoting the willingness of users in role $r_i$ adopting contagions posted or forwarded by users in role $r_j$; (b) User Role - Contagion Sentiment Interactions $\Omega_{s}^{role}(r_i, s_x)$, denoting the willingness of users in role $r_i$ adopting contagions of sentiment $s_x$; (c) Contagion Sentiment-Sentiment Interactions $\Lambda(s_x, s_y)$, with $s_x$ as the ordinate and $s_y$ as the abscissa, denoting the influence of contagions in sentiment $s_y$ has on the decision of whether to adopt contagions in sentiment $s_x$; (d) User Role - Contagion Category Interactions $\Omega_{categ.}^{role}(r_i, c_k)$, denoting the willingness of users in role $r_i$ adopting contagions of category $c_k$; (e) Contagion Category-Category Interactions $\Lambda_{categ.}(c_i, c_k)$, with $c_i$ as the ordinate and $c_k$ as the abscissa, denoting the influence of a contagion in category $c_k$ has on the decision of whether to adopt a contagion in category $c_i$. }
  \label{Interactions}
\end{figure*}

Throughout this section, we provide qualitative insights into the extent to which the interactions influence the adoption of contagions. Due to limits of space, we only show the interaction results under IAD w/ LDA-S. Please note that interaction results under IAD w/ LDA-S and IAD w/o S should be identical except sentiment-related interactions are further derived with LDA-S. After fitting the IAD w/ LDA-S model, $\Delta_{role}$, $\Omega_{t-s}^{role}$ and $\Lambda_{t-s}$ are obtained. The results can be simply processed to obtain $\Omega_{topic}^{role}$ and  $\Lambda_{topic}$, and then $\Lambda_{categ.}$ and $\Omega_{categ.}^{role}$ can be derived by the category classification method. In addition, we can also derive the sentiment-related interactions, i.e., $\Omega_{s}^{role}$ and $\Lambda_{s}$. Fig.~\ref{Interactions} shows the inferred interactions when $|t|=20$ and $K=2$. 

\begin{itemize}
\item \textbf{User Role-Role Interaction.}  In Fig.~\ref{Interactions}(a), it can be observed that authority users are more likely to adopt contagions forwarded by other authority users, rather than those from hub users or ordinary users, which indicates a status gradient on social roles seniority. Hub users would like to adopt contagions from ordinary users, rather than from hub users. The ordinary users prefer to adopt contagions from other ordinary users or hub users, while don't forward the contagions from authority users with a larger probability. The reason is that, although the number of adoptions from authoritative users is commonly large, the forwarding ratio, that is, (the number of forwarded contagions) / (the number of exposed contagions), from authoritative
users is not necessarily high, due to the large exposure count in the denominator. 
 
\item \textbf{User Role-Contagion Sentiment Interaction.} Fig.~\ref{Interactions}(b) shows that no matter what roles the users play, they are more likely to forward neutral and positive contagions than the negative contagions. It is intuitive that the users commonly prefer positive contagions to the negative ones. One possible reason why users prefer the neutral ones may be because the neutral contagions are not aggressive and less likely to conflict with other contagions, and thus easy to get accepted. From the perspective of the user roles, hub users are more likely to forward contagions than authority users and ordinary users. Particularly, the hub users like to forward the positive contagions very much. This finding is consistent with Fig. 5(d), where the hub users like to forward advertisements, as most of the advertisements are supposed to be positive.


\item \textbf{Contagion Sentiment-Sentiment Interaction.} Fig.~\ref{Interactions}(c) shows that when a negative contagion meets another negative contagion, it can promote each other's propagation. This phenomenon is reasonable as they probably express similar opinions and cooperate with each other to get spread. The neutral sentiment seldom suppresses itself, neither does the positive sentiment. The mutual suppression between positive contagions and negative contagions is very strong, and stronger than the suppressions between positive/negative and neutral contagions. It is intuitive as positive and negative contagions commonly express opposite opinions, and it is more likely for them to compete rather than cooperate with each other.


\item \textbf{User Role-Contagion Category Interaction.} Fig.~\ref{Interactions}(d) shows that authority users are more likely to adopt contagions on economy, news, and politics, and don't like to forward advertisements. One possible reason is that the authority users such as the news media give more attention to big events such as politics rather than small events in common life. On the contrary, ordinary users prefer contagions such as movies and food, and they don't like to forward advertisements and political contagions. Hub users tend to adopt contagions about advertisement and movies, and one possible reason is that they may be spam users.
%

\item \textbf{Contagion Category-Category Interaction.} Fig.~\ref{Interactions}(e) reveals how different categories of contagions compete or cooperate with each other to get propagated. It can be observed that generally the entries in the matrix are negative, which indicates that relationships between different categories are mainly competition. It validates the existing conclusion that attention is limited for individual users to adopt contagions~\cite{Competition}. It is obvious that the colors of the entries on the diagonal line are lighter, indicating that contagions affiliated to the same category are less likely to suppress each other. It is intuitive that when similar topics meet each other, they tend to cooperate to become hot topics and thus attract more attention, making themselves easier to get spread. It also shows that contagions belonging to food category are more likely to get adopted when simultaneously propagating with contagions belonging to other categories, i.e., the propagation of contagions on food are more likely to suppress the propagation of other contagions. In addition, contagions about constellation and life also attract a lot of attention. On the contrary, contagions belonging to categories such as the advertisement and news are less likely to suppress other contagions' propagation, revealing that commonly users don't like to forward them.

\end{itemize}

\section{Conclusion}
In this paper, a new information diffusion framework called IAD is proposed to analyze the users' behaviors on adopting a contagion, in consideration of the interactions involving users and contagions as a whole. With this framework, we can quantitatively study how these interactions would influence the propagation process. To efficiently learn the interactions, we use a generative process to infer user roles and a co-training method to classify the contagions into explicit categories. To involve sentiment factors into the user's forwarding behavior, we also propose a LDA-S model which are able to extract the sentiment distribution and topic distribution simultaneously from contagions. Experimental results on the large-scale Weibo dataset demonstrate that IAD methods can outperform the state-of-art baselines in terms of F1-score, accuracy and runtime. Moreover, IAD with sentiment can further improve the prediction performance of IAD without sentiments. Last but not least, various kinds of interactions can be obtained and interesting findings can be observed, which are useful to various domains such as viral marketing.

\section*{Acknowledgments}

This work was supported in part by State Key Development Program of Basic Research of China (No. 2013CB329605), the Natural Science Foundation of China (No. 61300014, 61672313), and NSF through grants IIS-1526499, IIS-1763325, and CNS-1626432, and DongGuan Innovative Research Team Program (No.201636000100038).




%

%

\begin{IEEEbiography}[{\includegraphics[width=1in,height=1.25in,clip,keepaspectratio]{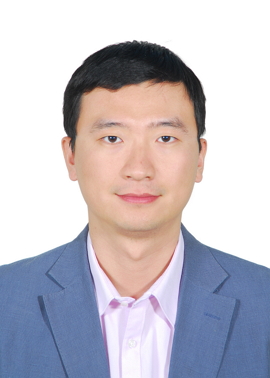}}]{Xi Zhang}
received the PhD degree in Computer Science from Tsinghua University. He is an
associate professor in Beijing University of Posts and Telecommunications, and is also the
vice director of Key Laboratory of Trustworthy Distributed Computing and Service, Ministry of Education, China. He was a visiting scholar
at the University of Illinois at Chicago. His research
interests include data mining and computer architecture. He is a member of IEEE.
\end{IEEEbiography}
\begin{IEEEbiography}[{\includegraphics[width=1in,height=1.25in,clip,keepaspectratio]{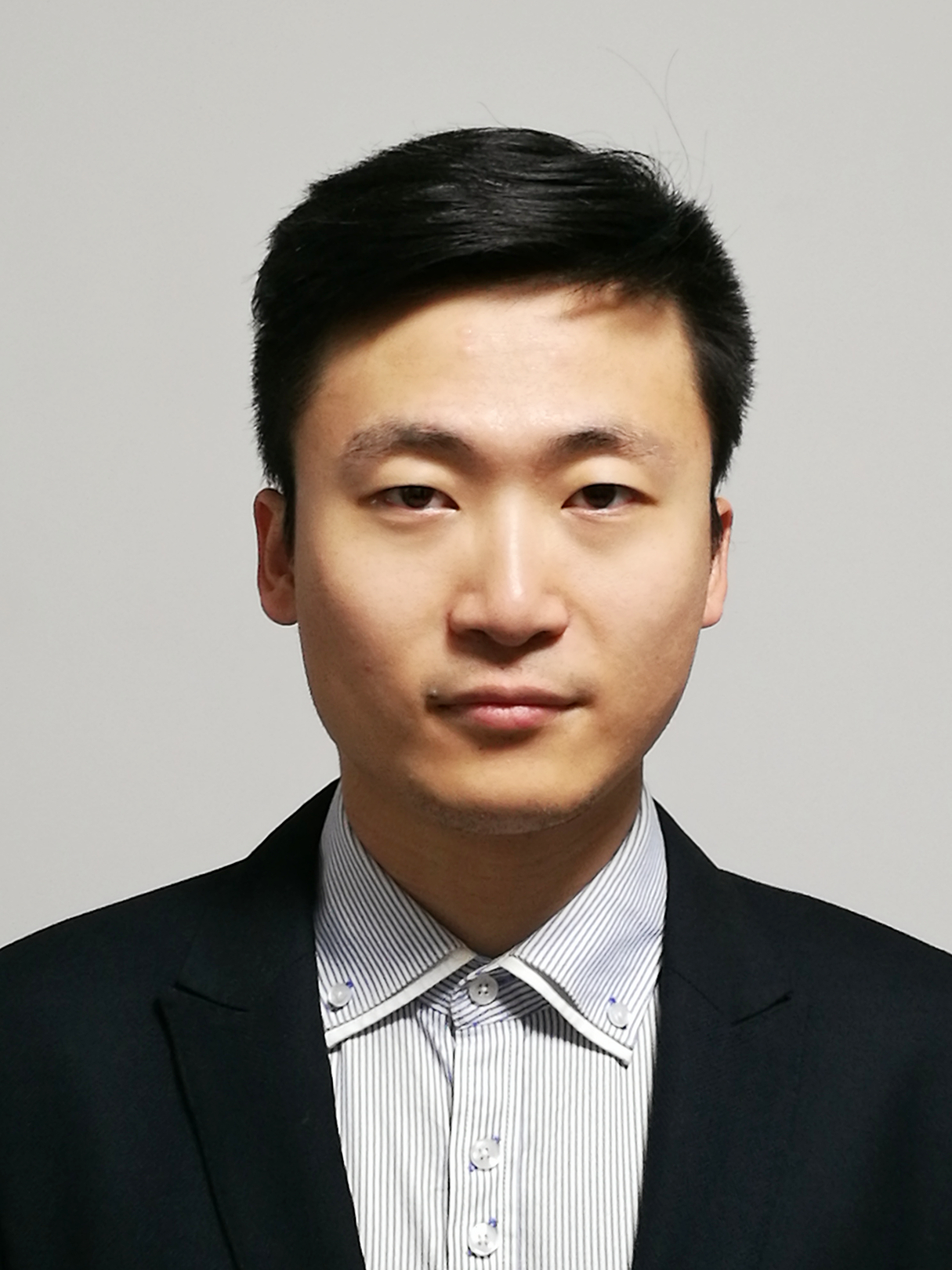}}]{Yuan Su}
received the master degree in Computer Science from Beijing University of Posts and Telecommunications in 2012. He is now a PhD student in Key Laboratory of Trustworthy Distributed Computing and Service (Beijing University of Posts and Telecommunications), Ministry of Education, China. He was a visiting student
at the University of Queensland from May 2015 to May 2016. His research
interests include data mining and social network analysis.
\end{IEEEbiography}
\begin{IEEEbiography}[{\includegraphics[width=1in,height=1.25in,clip,keepaspectratio]{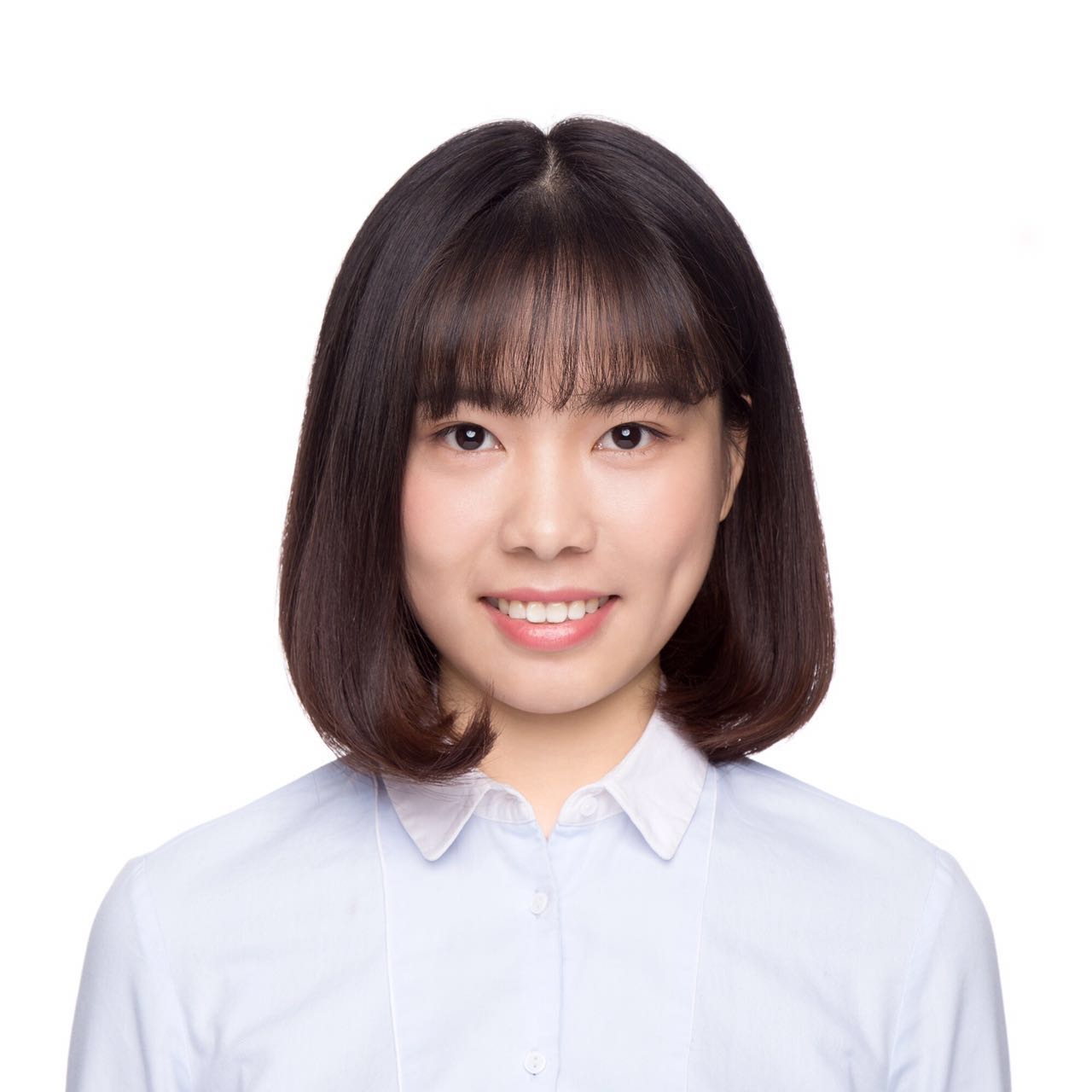}}]{Siyu Qu}
received the bachelor degree in Computer Science from Xidian University in 2012. She is now a master student in Key Laboratory of Trustworthy Distributed Computing and Service (Beijing University of Posts and Telecommunications), Ministry of Education, China. Her research
interests include data mining and machine learning.
\end{IEEEbiography}
\begin{IEEEbiography}[{\includegraphics[width=1in,height=1.25in,clip,keepaspectratio]{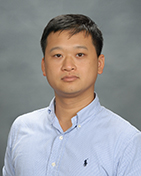}}]{Sihong Xie}
received the PhD Degree from the Department of Computer Science at University of Illinois at Chicago. He is an assistant professor in Department of Computer Science and Engineering, Lehigh University. His research
interests include big data, data mining, and machine learning.
\end{IEEEbiography}
\begin{IEEEbiography}[{\includegraphics[width=1in,height=1.25in,clip,keepaspectratio]{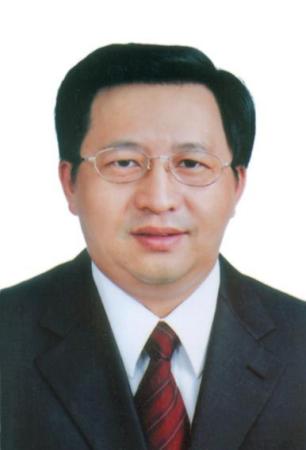}}]{Binxing Fang}
received his PhD
from Harbin Institute of Technology, China
in 1989. He is a member of the Chinese
Academy of Engineering and a professor
in School of Cyberspace Security at Beijing
University of Posts and Telecommunications. He is currently the chief scientist
of State Key Development Program of Basic
Research of China. His current interests
include big data and information
security.
\end{IEEEbiography}
\begin{IEEEbiography}[{\includegraphics[width=1in,height=1.25in,clip,keepaspectratio]{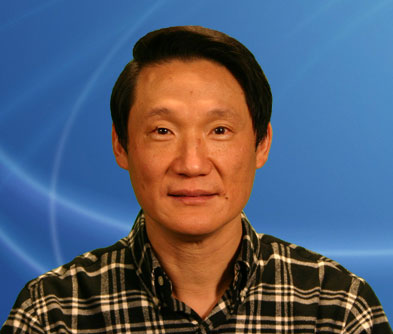}}]{Philip S. Yu}
received the PhD degree in electrical
engineering from Stanford University. He is a
distinguished professor in computer science at
the University of Illinois, Chicago, and is the Wexler
chair in Information Technology. His research
interests include big data, data mining, data
stream, database, and privacy. He was the editor-in-chief of the IEEE Transactions on Knowledge and
Data Engineering and the ACM Transactions on Knowledge
Discovery from Data. He received the
ACM SIGKDD 2016 Innovation Award, a Research Contributions Award
from the IEEE International Conference on Data Mining (2003), and a
Technical Achievement Award from the IEEE Computer Society (2013).
He is a fellow of the IEEE and the ACM.
\end{IEEEbiography}





\end{document}